%% file: TC_Transition.tex
\documentclass{jfm}

\usepackage{graphicx}
\usepackage[dvipsnames]{xcolor} 
\usepackage{multirow}
\usepackage{epstopdf, epsfig}
\usepackage[caption=false]{subfig} 
\usepackage{natbib}
\usepackage{amsmath}
\DeclareMathSymbol{\shortminus}{\mathbin}{AMSa}{"39}
\usepackage{amssymb}
\usepackage{comment}
\usepackage{booktabs}
\usepackage{url}
\usepackage{tikz}
\usepackage{textcomp}  
\usepackage{pstricks}
\usepackage{gensymb} 
\usepackage{import} 
\definecolor{mygreen}{rgb}{0,0.6,0}

\newcommand{\Reo}{\ensuremath{Re_o}} 
\newcommand{\Rei}{\ensuremath{Re_i}} 
\newcommand{\Reio}{\ensuremath{Re_{i,o}}} 
\newcommand{\Reic}{\ensuremath{Re_i^c}} 

\newcommand{\YellowSquare}{\raisebox{-0.0em}{\includegraphics[scale=0.45]{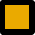}}}
\newcommand{\YellowTriangle}{\raisebox{-0.0em}{\includegraphics[scale=0.4]{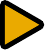}}}
\newcommand{\YellowDiamond}{\raisebox{-0.0em}{\includegraphics[scale=0.4]{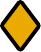}}}
\newcommand{\BlueCircle}{\raisebox{-0.0em}{\includegraphics[scale=0.4]{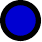}}}

\title{A novel subcritical transition to turbulence in Taylor-Couette flow with counter-rotating cylinders}
\shorttitle{Novel transition to turbulence in Taylor-Couette flow with counter-rotation}
\shortauthor{C.J. Crowley, M.C. Krygier, D. Borrero-Echeverry, R.O. Grigoriev, M.F. Schatz}

\author{Christopher J. Crowley,\aff{1}\textsuperscript{\corresp{\email{chris.crowley@gatech.edu}}} Michael C. Krygier,\aff{1}\\ Daniel Borrero-Echeverry,\aff{2} Roman O. Grigoriev,\aff{1}\\ and Michael F. Schatz\aff{1}}

\affiliation{\aff{1}Center for Nonlinear Science and School of Physics, Georgia Institute of Technology, \\ Atlanta, GA 30332, U.S.A.
\aff{2}Department of Physics, Willamette University, Salem, OR 97301, U.S.A.}

\date{\today}

\begin{document}

\maketitle

\begin{abstract}

The transition to turbulence in Taylor-Couette flow often occurs via a sequence of supercritical bifurcations to progressively more complex, yet stable, flows.  We describe a subcritical laminar-turbulent transition in the counter-rotating regime mediated by an unstable intermediate state in a system with an axial aspect ratio of $\Gamma=5.26$ and a radius ratio of $\eta=0.905$. In this regime, flow visualization experiments and numerical simulations indicate the intermediate state corresponds to an aperiodic flow featuring interpenetrating spirals. Furthermore, the reverse transition out of turbulence leads first to the same intermediate state, which is now stable, before returning to an azimuthally-symmetric laminar flow.  Time-resolved tomographic particle image velocimetry is used to characterize the experimental flows; these measurements compare favorably to direct numerical simulations with axial boundary conditions matching those of the experiments.

\end{abstract}


\section{Introduction} 
\label{sec:intro}

The transition to turbulence in many flows falls into two classes: subcritical transitions whereby the transition is directly from laminar flow to turbulence or supercritical transitions where the transition occurs through a sequence of intermediate stable flow states before ultimately ending in turbulence.   By describing well the growth of infinitesimal disturbances, linear stability theory has often enabled good predictions of the critical Reynolds numbers at which supercritical transitions occur.  By contrast, subcritical transitions result from the nonlinear growth of finite amplitude perturbations; therefore, linear stability analyses provide little insight into this type of transition.  Subcritical transitions exhibit hysteresis, in which the turbulent flow returns back to the laminar state at a Reynolds number that is lower than that for the transition from laminar flow to turbulence.

\begin{figure}
\vspace{0.15in}
	\centering
	\includegraphics[width=0.325\textwidth]{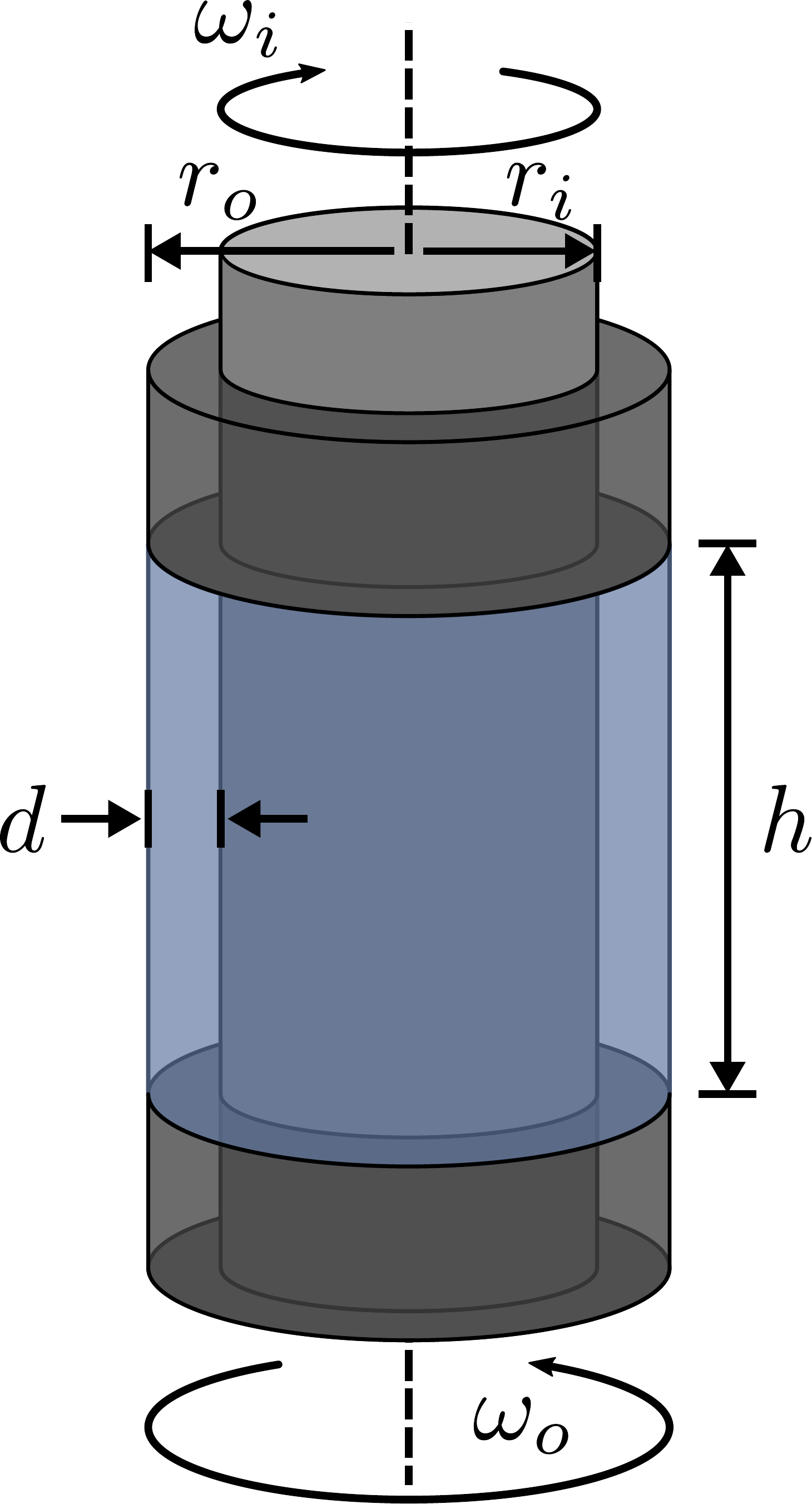}
	\caption{\label{fig:geometry} In Taylor-Couette flow, a fluid is confined between coaxial cylinders of radii $r_i$ and $ r_o$, which counter-rotate with angular velocities $\omega_{i}$ and $\omega_{o}$, respectively.  In the axial direction, the flow is bounded by two end caps that rotate with the outer cylinder and are separated by a distance $h$.  In the radial direction, the separation between the cylinders is $d = r_o - r_i$.  The flow is periodic in the azimuthal direction.}		
\end{figure}

Both super- and subcritical turbulent transitions can be observed in the flow between two independently rotating, coaxial cylinders, or Taylor-Couette flow (TCF) (see Figure \ref{fig:geometry}).    TCF can be uniquely characterized by four nondimensional parameters.  Two parameters characterize the geometry of the system: the radius ratio $\eta=r_i/r_o$, where $r_i$ and $r_o$ are the radii of the inner and outer cylinders, respectively, and the aspect ratio $\Gamma=h/d$, where $d = r_o-r_i$ is the radial separation distance between the cylinders and $h$ is the axial height of the flow domain.  The other two parameters, the inner and outer Reynolds numbers \Reio\, describe the cylinders' rotation rates and are given by
\begin{equation}
	\Reio=\frac{r_{i,o}\,\omega_{i,o}\,d}{\nu},
\end{equation}
\noindent where $\nu$ is the kinematic viscosity of the fluid and $\omega_{i,o}$ are the angular velocities of the  inner and outer cylinders, respectively. By convention \Rei\ is always taken to be positive, whereas \Reo\ is positive when the cylinders are co-rotating and negative when they are counter-rotating.

Numerous studies of supercritical transitions to turbulence in TCF have been published (see, e.g., \citet{Coles1965, Andereck1986, Tagg1994, Meseguer2009a}).  Experimental studies mostly focus on geometries with $\Gamma \gg 1$ to reduce the influence of the axial boundaries on the flow and investigate transition at fixed \Reo\ while quasi-statically increasing \Rei.  In this parameter regime, TCF exhibits a multitude of stable non-turbulent flow states for different rotation rates \citep{Coles1965,Andereck1986}.   As \Rei\ is increased, each new transition typically yields a flow of increased complexity until the flow eventually becomes turbulent. 

The subcritical transition to turbulence in TCF has a long history \citep{Couette1890, Mallock1896, Wendt1933, Taylor1936, Taylor1936a, SchultzGrunow1959,Coles1965,VanAtta1966,Andereck1986,Prigent2005}. Recently, there has been renewed interest in this regime as a testbed for ideas explaining the subcritical transition to turbulence from the viewpoint of dynamical systems theory \citep{Meseguer2009b,Borrero2010,Avila2013b,Maretzke2014,Lopez2016} and as a model for understanding the source of the enhanced angular momentum transport observed in astrophysical disks, which is purportedly caused by turbulence despite the predicted stability of these flows \citep{Richard1999,Ji2006,Paoletti2011,Burin2012,Edlund2014}. In these studies, which were mostly conducted with $\Gamma \gg 1$, transition is typically observed to take place directly from the laminar base flow. 

The subcritical transition to turbulence is commonly associated with wall-bounded shear-driven flows in channels, pipes and boundary layers in which two general transition scenarios have been identified: (1) an ``amplification" scenario involving transient non-turbulent modes \citep{Reshotko1976} and (2) a ``bypass" transition \citep{Morkovin1985}.  In the amplification scenario, the transition begins with the appearance of a weak structured flow (e.g., Tollmien-Schlichting (T-S) waves in channel flow), which undergoes amplification by linear mechanisms until it is sufficiently large that nonlinearity takes over; eventually, the flow breaks up into turbulence. In this scenario, the unstable structured flow acts as a transient intermediary that triggers turbulence.  Operationally, the amplification transition is observed by imposing carefully controlled disturbances in flows where the ambient free-stream turbulent intensities are sufficiently low \citep{Nishioka1975}. In contrast, in the bypass transition scenario, the free-stream turbulent intensity is sufficiently large that non-turbulent structured flows are not observed (are ``bypassed'') and, for sufficiently large $Re$, the transition occurs from featureless laminar flow directly to disordered turbulent flow.  Earlier studies of the subcritical transition to turbulence in TCF, which were mostly conducted at large aspect ratios, showed that this transition typically occurs directly from the laminar base flow, i.e., via the bypass mechanism.

Here we report a subcritical transition to turbulence in counter-rotating TCF that has some features in common with the amplification scenario found in prior work on other wall-bounded shear flows.  In our TCF studies, a non-turbulent intermediary flow plays a central role in the transition. This aperiodic flow features interpenetrating spirals (IPS) with opposite helicity much akin to those that have been discovered by \citet{Andereck1986} at large $\Gamma$ and further investigated by \citet{Coughlin1996}.  At a moderate value of $\Gamma$ = 5.26, we find that IPS appear transiently in the transition from laminar to turbulent flow as \Rei\ is increased at fixed \Reo.  If \Rei\ is then decreased, the turbulent flow transitions to \textit{stable} IPS, which persist over a range of \Rei. As \Rei\ is decreased further, stable IPS eventually transition back to laminar flow. 

It is important to note that both super- and subcritical transitions have been examined in earlier studies of TCF with small-to-moderate aspect ratios ($\Gamma \lesssim 5$). A rich variety of phenomena have been observed, including a plethora of asymmetric states from symmetric end cap forcing \citep{Tavener1991}, a sequence of period doubling bifurcations \citep{Pfister1988}, and quasiperiodic dynamics with three frequencies \citep{Lopez2003}.  These examples and many others from prior work describe transitions to non-turbulent flows, distinct from the transition to turbulence reported here.

In Section \ref{sec:setup}, we outline briefly the experimental and numerical methods used in this study.  Then, in Section \ref{sec:results}, we discuss the transitions between flow states with a particular emphasis on the role of the IPS.  In Section \ref{sec:discussion}, we discuss the implications of this discovery for understanding of the subcritical transition to turbulence in TCF and, more generally, in wall-bounded shear flows; we conclude in Section \ref{sec:conclusion}.

\section{Methods} 
\label{sec:setup}

Our TCF apparatus with $\eta = 0.905$ was composed of a glass outer cylinder with a radius of $r_o=80.03\pm0.02$\;mm  and a brass inner cylinder of radius  $r_i=72.39\pm0.01$\;mm with a black powder coat to enhance optical contrast in flow visualization studies.  The aspect ratio, $\Gamma = 5.26$, was set by two end caps, separated axially by $h = 40.2\pm0.05$\;mm and attached to rotate with the outer cylinder. The cylinders were driven by stepper motors; to reduce vibration and to ensure uniform cylinder rotation, timing belts connected the cylinders to the motors, which were mounted separately from the TCF apparatus.  Additionally, a transmission with a gear ratio of 28:1 was used with the inner cylinder stepper motor to increase the resolution in \Rei. While the cylinders were rotating, the rate of temperature variations in the flow was kept below $0.5\;{}^{\circ}\text{C}$ throughout the duration of the experiments by surrounding the outer cylinder with a liquid bath. With these measures, the total systematic uncertainty for \Rei\ and \Reo\ was below 1\;\%.

The flow was characterized using rheoscopic flow visualization. In some studies, the working fluid was water mixed with Kalliroscope \citep{Matisse1984} at a concentration of 0.3\;\%  by volume and had a kinematic viscosity of $\nu=1.01\;\textrm{mm}^2/\textrm{s}$ at 20\;${}^{\circ}$C. Other studies were carried out using a mixture of water and a stearic-acid-based rheoscopic fluid \citep{Borrero2018} at a concentration of 5\;\% by volume with kinematic viscosity $\nu=1.03\;\textrm{mm}^2/\textrm{s}$ at 20\;${}^{\circ}$C. The flows were illuminated using fluorescent lights and imaged using a single Microsoft LifeCam HD webcam oriented perpendicular to the flow domain and connected via a USB interface to a computer. The resulting digital images were analyzed using a custom Matlab script to identify qualitative changes in the flow as a function of Reynolds number as an indicator of flow transition. For each image, the script counted the total number of pixels with an intensity above a fixed threshold; different flow states exhibited different, easily distinguishable pixel counts.

Tomographic particle image velocimetry (tomo PIV) \citep{Elsinga2006} was also used to perform flow measurements.   In tomo PIV, particles suspended in the flow are imaged simultaneously by multiple cameras at different viewing angles, and the images are used to reconstruct the light intensity distribution in a 3-D flow volume; 3-D cross-correlation of distributions reconstructed at different times enables determination of 3-D velocity fields throughout a flow volume with an approximate size of $d$ radially, $0.75 h$ axially, and $2\pi r_o/10$ azimuthally.  Custom-made, density-matched polyester particles  (25\:\textmu{}m to 32\:\textmu{}m in diameter) were doped with Rhodamine 6G and suspended in the flow. The particles were illuminated with a Quantronix 527/DP-H Q-switched Nd:YLF laser.  Fluorescent light emitted from the particles was collected by four Vision Research Phantom V210 high speed cameras synchronized with the laser illumination.  Each camera was fitted with a 105 mm Nikon Nikkor fixed focal length lens attached via a Scheimpflug adapter (LaVision Inc.).  A low pass optical filter (Semrock BLP01-532R-25) on each camera lens attenuated, by a factor of $10^7$, the scattered 527\;nm wavelength laser illumination and passed, with 80\% efficiency, fluorescent light at wavelengths $>$532\;nm.  The images were then analyzed using LaVision Inc.'s DaVis tomographic PIV software package.  To reduce optical distortion from the outer cylinder's curved surfaces, the index of refraction of both the working fluid and the bath liquid were matched to the index of refraction of the glass outer cylinder.  Index matching of the working fluid was achieved by using an ammonium thiocyanate solution prepared with a specific gravity of 1.13 and a kinematic viscosity of $\nu=1.37\;\textrm{mm}^2/\textrm{s}$ at 23\;${}^{\circ}$C \citep{Borrero2016a}.  A small amount of ascorbic acid was added to the ammonium thiocyanate solution to mitigate reaction with trace metals \citep{Sommeria1991}.  Index matching of the bath liquid was achieved by a binary mixture of two mineral oils with a 68.8 \% heavy viscosity oil (McMaster-Carr part no. 3190K632) to 31.2 \% light viscosity oil (McMaster-Carr part no. 3190K629) ratio. Further details about the implementation of tomo PIV measurements in our TCF apparatus are reported elsewhere \citep{BorreroThesis}.

Fully resolved direct numerical simulations (DNS) of TCF were conducted using the code developed by M. Avila and his collaborators \citep{Avila2008,Mercader2010,Avila2012}. This code uses a pseudospectral scheme to solve the Navier-Stokes equation in cylindrical coordinates $\left(r,\theta,z\right)$ subject to physical (no-slip) boundary conditions at the surface of the rotating concentric cylinders and top and bottom end caps.  The geometry of the simulation was chosen to match that of the experimental apparatus.  The simulations used $N_r=20$ Chebyshev modes in the radial direction, $N_z=100$ Chebyshev modes in the axial direction, and $N_\theta=1280$ Fourier modes in the azimuthal direction, so that the velocity field $\bf{v}$ at a point $(r,\theta,z)$ and time $t$ is given by
\begin{align}
{\bf v}(r,\theta,z,t)= \mathrm{Re} \sum_{k=0}^{N_r}\sum_{n=0}^{N_z}
\sum_{m=0}^{N_\theta/2}{\bf V}^{knm}(t)T_k(x)T_n(y)e^{im\theta},
\end{align}
where $x=(2r-r_i-r_o)/d$, $y=2z/h-1$ (where $0<z<h$), and $T_n(\cdot)$ is the Chebyshev polynomial of order $n$. All experimental and numerical results are nondimensionalized in terms of a characteristic length scale $d=r_o-r_i = 7.64$\;mm (the annular gap width) and a characteristic (viscous) time scale  $\tau = d^2/\nu = 56.7$\;s.

To quantify flow fields in both simulations and experiments, the perturbation flow field 
\begin{equation}
    \widetilde{\mathbf{v}}(t) = \mathbf{v}(t)- \mathbf{v}^{\textrm{lam}},
\end{equation}
characterizes the deviation of the full flow $\mathbf{v}(t)$ from an axially symmetric laminar flow $\mathbf{v}^{\textrm{lam}}$ computed numerically at the same Reynolds numbers.  The numerically computed $\mathbf{v}^{\textrm{lam}}$ was used to compute the perturbation flow field for both simulations and experiments since the laminar flow is unstable for some \Rei\ considered in this study, and, therefore, unobservable in the laboratory experiments.

\section{Results}
\label{sec:results}

We first briefly describe a coarse experimental exploration of laminar-turbulent and turbulent-laminar transitions over a range of \Reo\ for cylinders that counter-rotate (\mbox{$Re_i>0$} and $Re_o < 0$).  We then focus on the case of $Re_o =-1000$ and examine in detail the transitions associated with increasing and decreasing \Rei\ using both laboratory experiments and numerical simulations.

\subsection{Laminar-Turbulent Transition: Dependence on \Reo} 
\label{sec:manyReo}
To coarsely map out the transition boundaries for TCF in the geometry studied here, we performed flow visualization experiments by first spinning up the outer cylinder from rest (with the inner cylinder stationary) to a specific value of \Reo; then, with \Reo\ held constant, \Rei\ was increased in steps of $\Delta\Rei = 10$ by slowly stepping up the rotation rate of the inner cylinder until a qualitative change in the flow was observed. We waited a time interval of $3.2\,\tau$ between steps to ensure that the flow had reached equilibrium. The turbulent-laminar  transition boundary at the same \Reo\ was then determined by starting in the turbulent regime and slowing the inner cylinder down by $\Delta\Rei = 10$ every $3.2\,\tau$ until the flow was observed to be in the laminar state.  The experiments were repeated for different fixed values of \Reo. 

\begin{figure}
	\centering	
	\def\svgwidth{1\textwidth}
    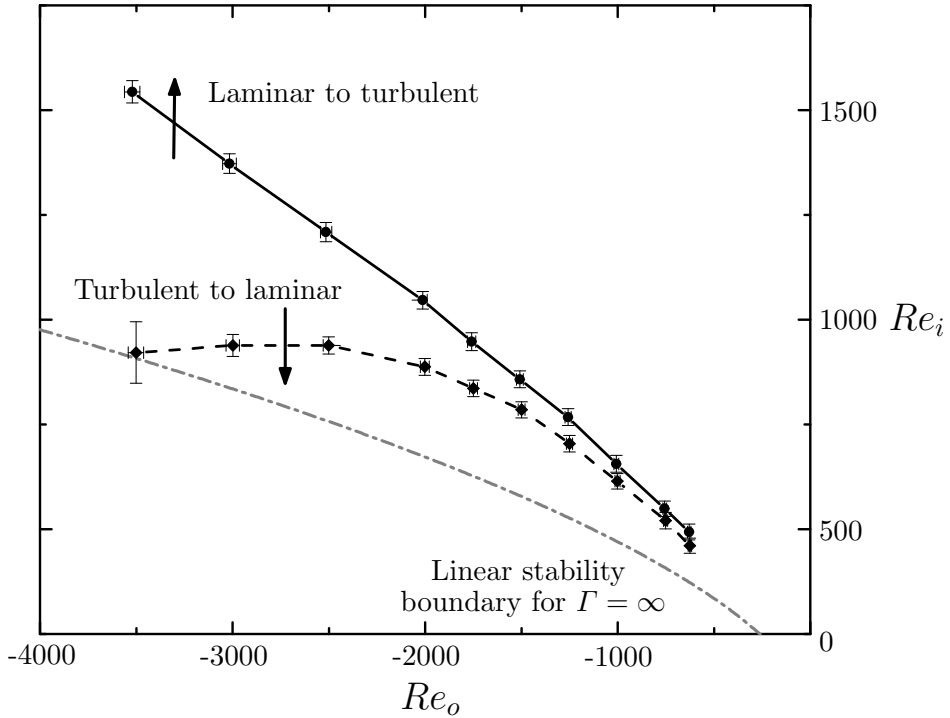
    \caption{\label{fig:full_transition} Phase diagram illustrating the hysteretic laminar-turbulent transition in experiments on counter-rotating ($Re_o < 0$ and $Re_i> 0$) Taylor-Couette flow with $\Gamma=5.26$ and $\eta=0.905$.   The diagram indicates transitions observed in experiments where \Rei\ was increased or decreased quasi-statically while keeping \Reo\ fixed.   The black solid and dashed lines, drawn to guide the eye, indicate the transition boundaries from laminar flow to turbulence and from turbulence to laminar flow, respectively. The gray dash-dotted line represents the marginal stability curve for TCF at $\eta = 0.9$ for $\Gamma=\infty$ \citep{Esser1996}. }
\end{figure}

The experimental studies revealed instability of the azimuthally symmetric smooth laminar flow always leads to turbulence over a range of \Reo\ from $-3500$ to $-500$ (Figure \ref{fig:full_transition}).  The transition back to laminar flow was always observed to be hysteretic; the range in \Rei\ over which hysteresis occurs increases as the magnitude of \Reo\ increases. Our results indicate that transition from laminar flow is suppressed by the moderate aspect ratio of our apparatus, i.e., for fixed \Reo, the transition occurs at \Rei\ larger than that predicted by linear stability analysis with $\Gamma=\infty$ (gray line in Figure \ref{fig:full_transition}).  This observation is consistent with earlier experiments at larger values of $\Gamma$ (and somewhat smaller  values of $\eta$) where, like our studies, the endcaps rotated with the outer cylinder \citep{Hamill1995}.  In that work, the delay of laminar flow transition was found to increase with decreasing $\Gamma$, most likely due to the end-wall effects (e.g. dissipation and Ekman pumping) that become more pronounced as $\Gamma$ decreases.

\subsection{Flow Transitions at $Re_o =-1000$} 
\label{sec:oneReo}

A detailed experimental and numerical investigation at fixed $Re_o =-1000$ led to the observation of an intermediate state that plays an important role in the laminar-turbulent transition. The transition from turbulence to laminar flow was found to involve an aperiodic stable intermediate state (interpenetrating spirals) that persists over a range of \Rei. Moreover, IPS were found to appear -- albeit transiently -- during the transition from laminar flow to turbulence. Transitions between different flow states are described in detail below.

\begin{table}
   \centering
    \begin{tabular}{ c c c c }
        \textbf{Transition} & \textbf{Experiment} & \textbf{Noiseless DNS} \\
        Laminar $\rightarrow$ Turbulence & 643 $\pm$ 2 &  675 $\pm$ 5 \\  
        Turbulence $\rightarrow$ IPS & 625 $\pm$ 3.6 & 623.5 $\pm$ 0.5 \\
        IPS $\rightarrow$ Turbulence & 631 $\pm$ 3.7 & 630.5 $\pm$ 0.5 \\       
        IPS $\rightarrow$ Laminar & 617 $\pm$ 1 & 617.5 $\pm$ 0.5 \\      
    \end{tabular}
			    \caption{\label{values} The inner cylinder Reynolds numbers for flow transitions are shown for both laboratory experiments and numerical simulations at $Re_o =-1000$. Uncertainty values from the experiment reflect the systematic uncertainties associated with the measurement of $Re$ as well as repeatability of the transition while the uncertainty values from the noiseless DNS reflect the resolution with which \Rei\ was investigated.}
\end{table}

\subsubsection{Transitions in Laboratory Experiments} 
\label{sec:results_lam_turb}

Transitions were determined in flow visualization studies by first spinning up the outer cylinder to $Re_o =-1000$ (with the inner cylinder at rest), and then increasing the inner cylinder's counter-rotation in steps of $\Delta \Rei = 0.5$ every $3.2\,\tau$, until the flow became turbulent. Subsequently, beginning from the turbulent state, \Rei\ was decreased at the same rate as before until the flow returned to the laminar state. No observable shifts in the transition boundaries were found when incrementing or decrementing \Rei\ in steps of $\Delta \Rei = 0.25$ separated in time by $10.7\,\tau$.    

\begin{figure}
	\centering
	\subfloat[]{\includegraphics[width=0.32\columnwidth]{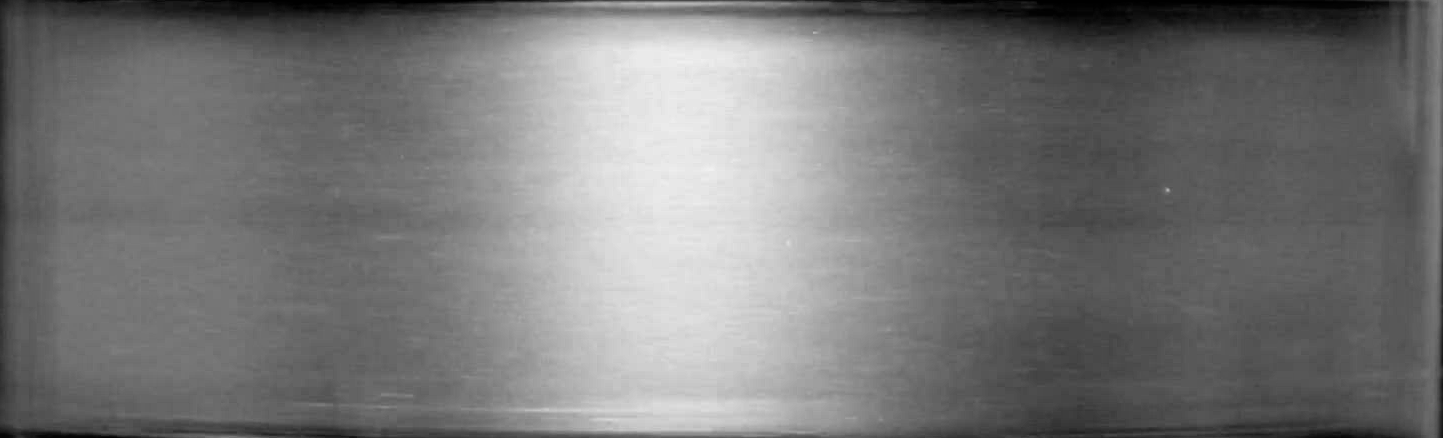}}\hspace{1mm}
	\subfloat[]{\includegraphics[width=0.32\columnwidth]{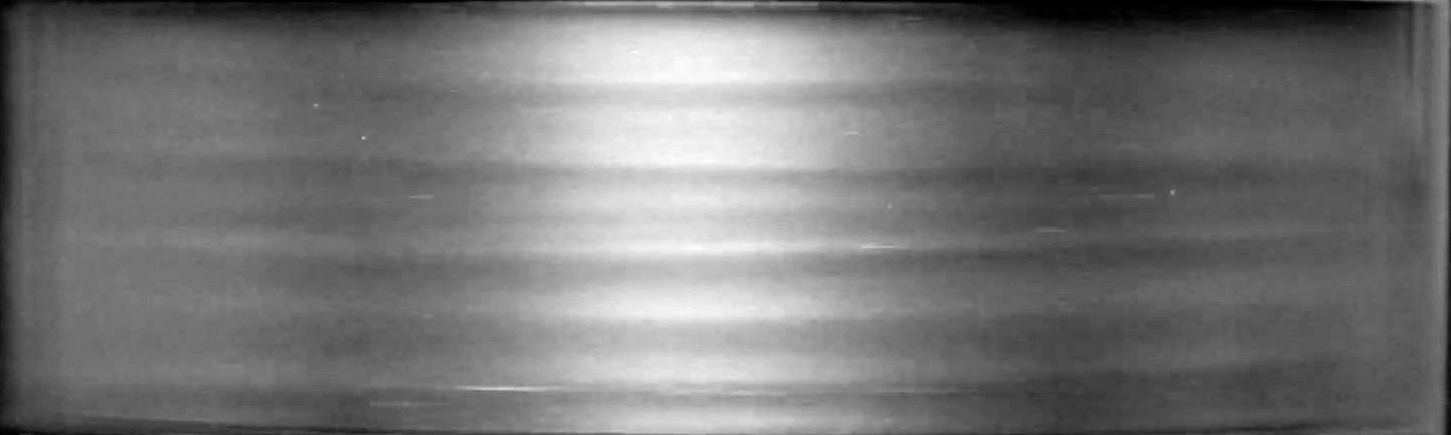}}\hspace{1mm}
	\subfloat[]{\includegraphics[width=0.32\columnwidth]{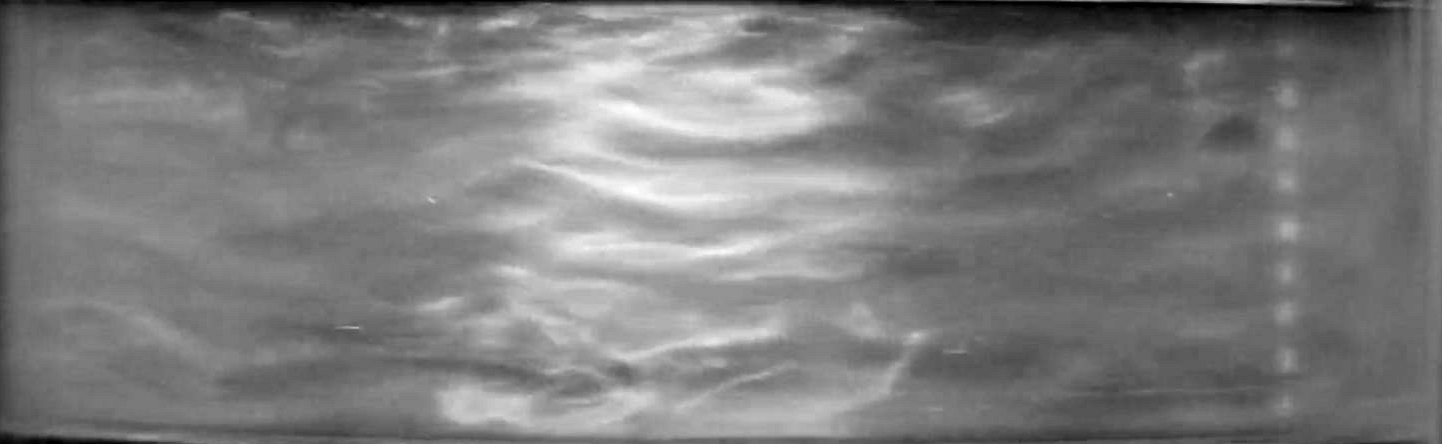}}
	\caption{\label{exp_turb2lam} Evolution of the flow during laminar to turbulent transition in experiments at  $Re_o =-1000$  and $Re_i =643$.  	The sequence of snapshots shows (a) the initial laminar flow, (b) transient interpenetrating spirals, and (c) persistent intermittent turbulence.  }
\end{figure}
With the flow starting in a laminar state, laboratory experiments exhibit a laminar-turbulent transition at $Re_i =643$ with a total uncertainty in \Rei\ of $\pm2$. Repeated measurements demonstrate the transition can be determined to a resolution of $\pm0.13$ (i.e. 0.02\;\%), as constrained by the mechanical limits of the motor and transmission driving the inner cylinder; in other words, from laminar flow just below threshold (cf. Figure \ref{exp_turb2lam}(a)), a single increment of $\Delta\Rei = 0.13$ reproducibly results in turbulence. At onset (with \Rei\ fixed), the structure of the flow changes slowly at first; very weak interpenetrating spirals gradually become discernible and grow slowly in amplitude with time (cf. Figure \ref{exp_turb2lam}(b)). Then, abruptly, the spirals break up and spatiotemporally intermittent turbulence develops on top of an IPS-like background flow and persists (cf. Figure \ref{exp_turb2lam}(c)). The interval of time over which the flow resembles IPS before transitioning to turbulence was different each time the experiment was performed and this interval decreased with an increase in the increment size of $\Delta \Rei$.  If \Rei\ is increased stepwise (with a fixed time interval of $3.2\,\tau$ between each step), the transition \Rei\ is unchanged for increments of $\Delta\Rei < 1$; the transition \Rei\ is observed to decrease for increments of $\Delta\Rei >1$.

Starting from turbulent flow, decreasing \Rei\ reveals a transition to stable IPS at $Re_i =625\pm3.6$. IPS  were observed to be weakly chaotic (i.e., having a broad-band temporal spectrum) over a range of \Rei\ and persist for as long as $3.8\times10^{3}\,\tau$ (two and a half days, after which time the experiments were ended). From stable IPS, increasing \Rei\ leads to a transition back to intermittent turbulence at $Re_i = 631\pm3.7$; decreasing \Rei\ leads instead to a transition to the axisymmetric laminar state at $Re_i = 617\pm1$. It should be noted that the values of \Rei\ at which various transitions are observed (see Table \ref{values}) depend on disturbances of two qualitatively different types: (a)  disturbances associated with a discrete change of \Rei\ and (b) other types of disturbances (e.g., the cylinders not being perfectly round or coaxial, the deviation in their angular velocity from a constant, etc.). All of these are disturbances of a finite, though likely small, amplitude.

\begin{figure}
	\centering	
	\def\svgwidth{1\textwidth}
    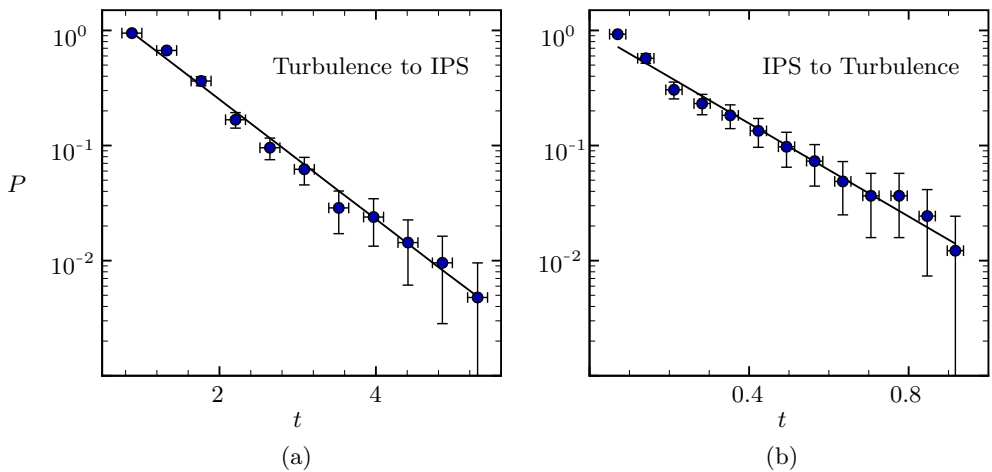
	\caption{\label{fig:stats} Transition probability $P(t)$. After an abrupt change in \Rei, the transition time for turbulence to either (a) disappear or (b) first appear is exponentially distributed.  $P(t)$ indicates the fraction of experimental trials where either (a) turbulence still persists after changing \Rei\ from 640 to 623 or (b) turbulence has not yet appeared after  changing \Rei\ from 623 to 640.
}
\end{figure}

A series of quenching experiments \citep{bottin1998,Prigent2005,Peixinho2006,Borrero2010} were carried out to characterize further the hysteretic transition between IPS and turbulence.  To probe the transition from turbulence to IPS, a turbulent state at $Re_i = 640$ was first established and monitored for $5.3\,\tau$; the inner cylinder rotation was then rapidly (in approximately $4.2\times10^{-3}\,\tau$) reduced to $Re_i = 623$. The time interval between the reduction in \Rei\ and the disappearance of turbulence was recorded. Similarly, to probe the transition from IPS to turbulence, IPS at $Re_i = 623$ was monitored for  $5.3\,\tau$; then, the inner cylinder rotation was rapidly (in approximately $4.2\times10^{-3}\,\tau$) increased to $Re_i = 640$ and the time interval between the increase in \Rei\ and the first appearance of turbulence was recorded. Figure \ref{fig:stats} summarizes the results from 250 experiments performing the same cycle of transitions between turbulence and IPS; the data indicate a clear exponential distribution of intervals between the time when the inner cylinder rotation rate is changed and the time for turbulence either completely disappears (cf. Figure \ref{fig:stats} (a)) or first appears (cf. Figure \ref{fig:stats} (b)). The exponential distribution suggests that both transitions describe a memoryless Poisson process, with a chaotic attractor at the initial \Rei\ becoming a chaotic repeller at the final \Rei\ \citep{Kadanoff1984,Kantz1985}. Similar distributions of transition lifetimes were found in relaminarization studies of high-aspect-ratio TCF \citep{Borrero2010}.

Quenching experiments were also performed for the transition from stable IPS to laminar flow; however, due to the discreteness with which \Rei\ could be varied in experiment, we could not find a final \Rei\ for which a meaningful distribution of lifetimes could be observed.

\begin{figure}
	\centering
	\subfloat[]{\includegraphics[width=0.325\columnwidth]{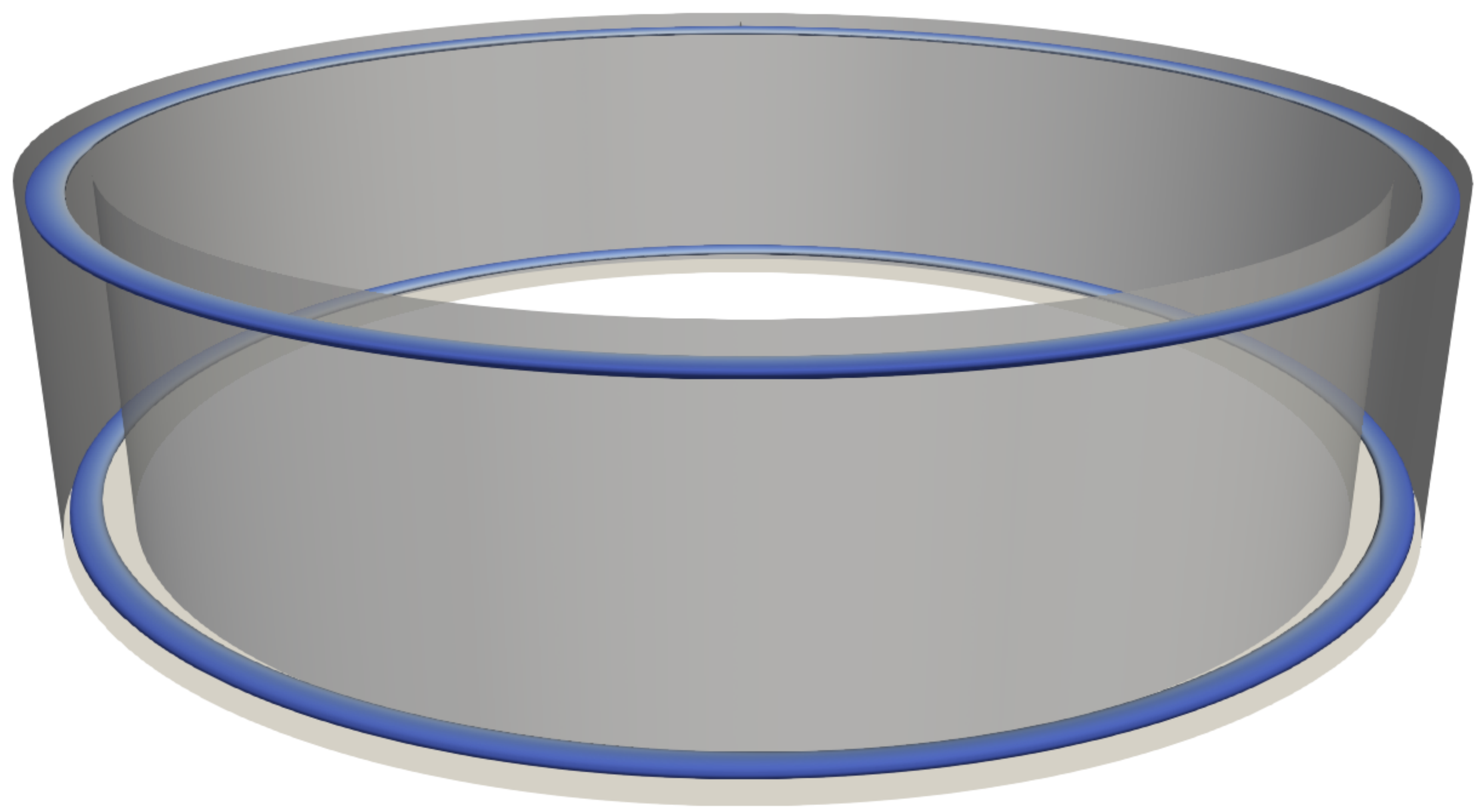}}\hspace{0mm}
	\subfloat[]{\includegraphics[width=0.325\columnwidth]{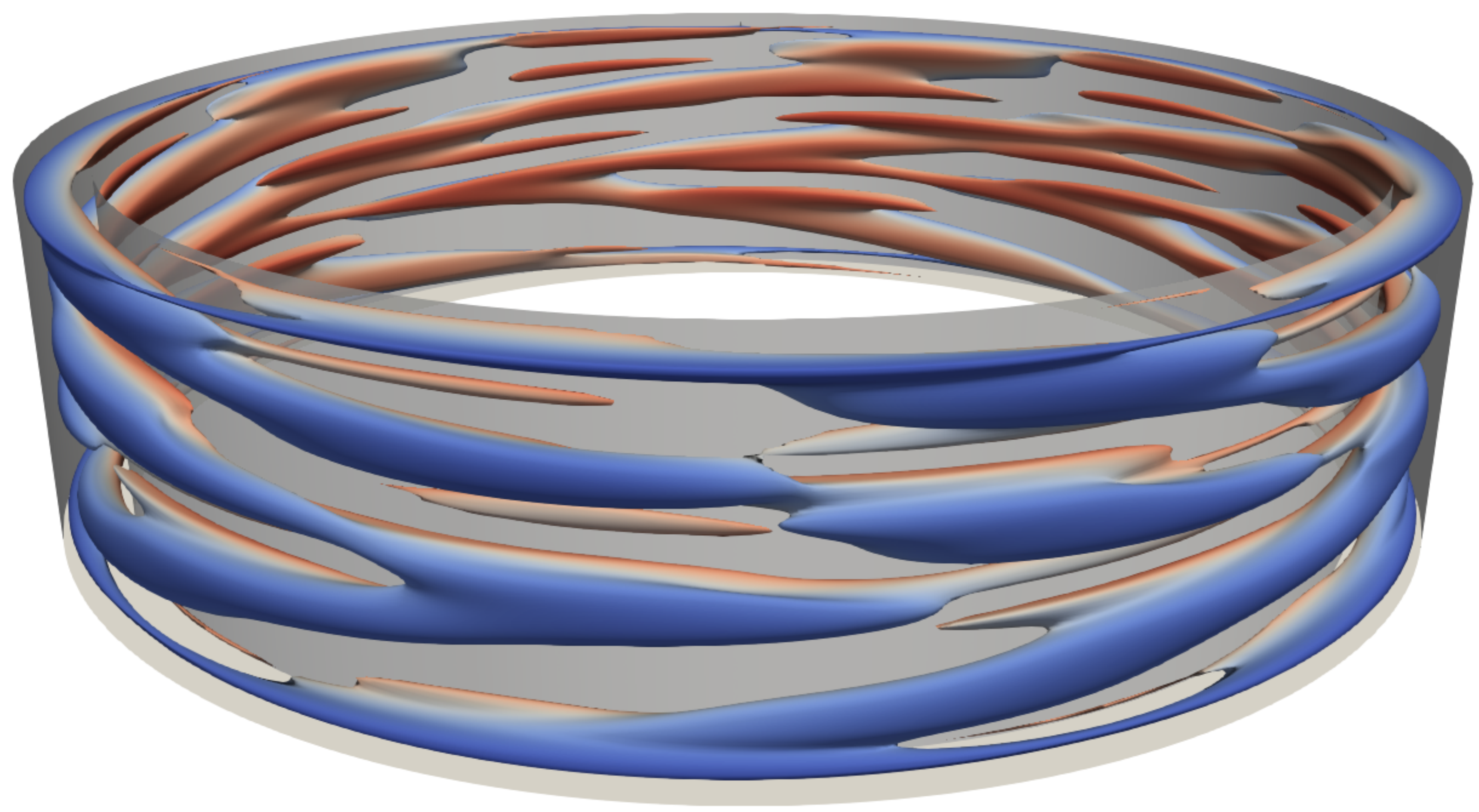}}\hspace{0mm}
	\subfloat[]{\includegraphics[width=0.325\columnwidth]{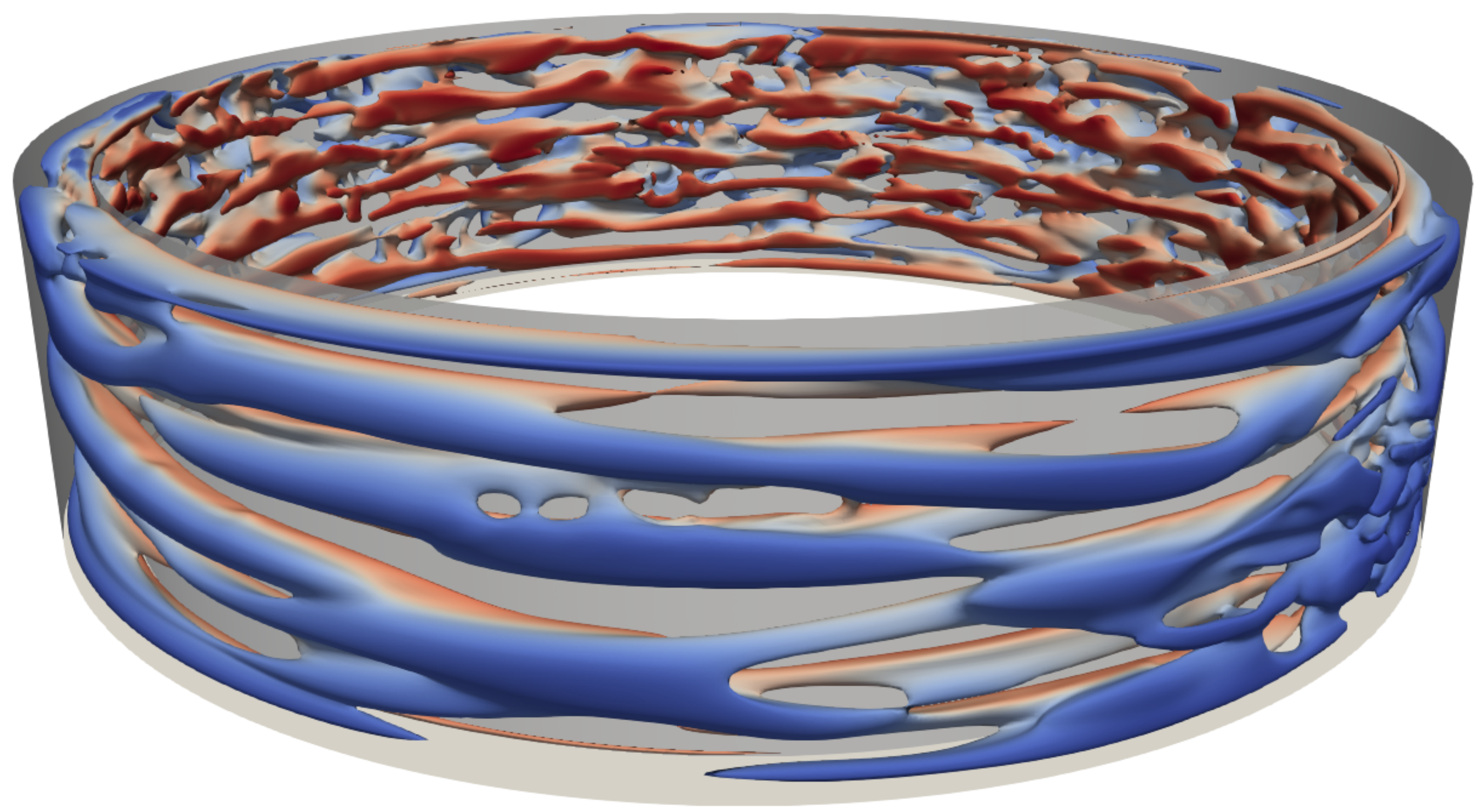}}
	\caption{\label{flow_fields_transition} Evolution of the flow during the laminar to turbulent transition in numerical simulations at $Re_o =-1000$  and $Re_i = 645$. (a) A laminar flow becomes unstable, transitioning first to (b) interpenetrating spirals and, eventually, (c) turbulence. Each panel shows isosurfaces of the radial velocity component; the color indicates the corresponding azimuthal velocity component. Red (blue) indicates flow in the same direction as the inner (outer) cylinder rotation. }
\end{figure}

\subsubsection{Transitions in Numerical Simulations} 
\label{sec:results_turb_ips}

Numerical simulations were used to determine linear stability of the steady axisymmetric laminar flow ${\bf v}^{\mathrm{lam}}$. This flow was generated at $Re_o =-1000$ and different fixed \Rei\ by keeping only the $m=0$ azimuthal Fourier mode and evolving the state until it stopped changing. The azimuthal symmetry of this flow was then broken by perturbing the $m=1$ Fourier mode (with the nonlinear term generating disturbances for all $m\ne 0$). Specifically, a random Gaussian noise with standard deviation $\sigma=10^{-8}$ was added to the coefficient of each of the spectral modes ${\bf V}_i^{kn1}\vert_{t=0}$, $i=r,\theta,z$ (note this is a very small perturbation since ${\bf V}^{kn0}_\theta=O(Re_o)$). Evolving the perturbed flow, we found that the perturbation decays (the laminar state is linearly stable) for $Re_i<Re_i^c = 675 \pm 5$ and grows, resulting in a transition to turbulence, for $Re_i>Re_i^c$.

Since the laminar flow undergoes transition to turbulence in experiment at a notably lower \Rei\ than the linear stability threshold \Reic, an investigation of its stability to finite amplitude disturbances was performed. Qualitatively, we find that, for $Re_i\ge 634$, finite amplitude perturbations lead to destabilization of the laminar state (cf. Figure \ref{flow_fields_transition}(a)), giving rise to IPS with an amplitude that grows and saturates temporarily (cf. Figure \ref{flow_fields_transition}(b)). Ultimately the IPS gives way to spatiotemporally intermittent turbulence (cf. Figure \ref{flow_fields_transition}(c)), just as in the experiment. The same transition sequence was found to occur for initial disturbances with different magnitudes and spatial profiles. 

\begin{figure}
	\centering
	\subfloat[]{\includegraphics[scale=0.139]{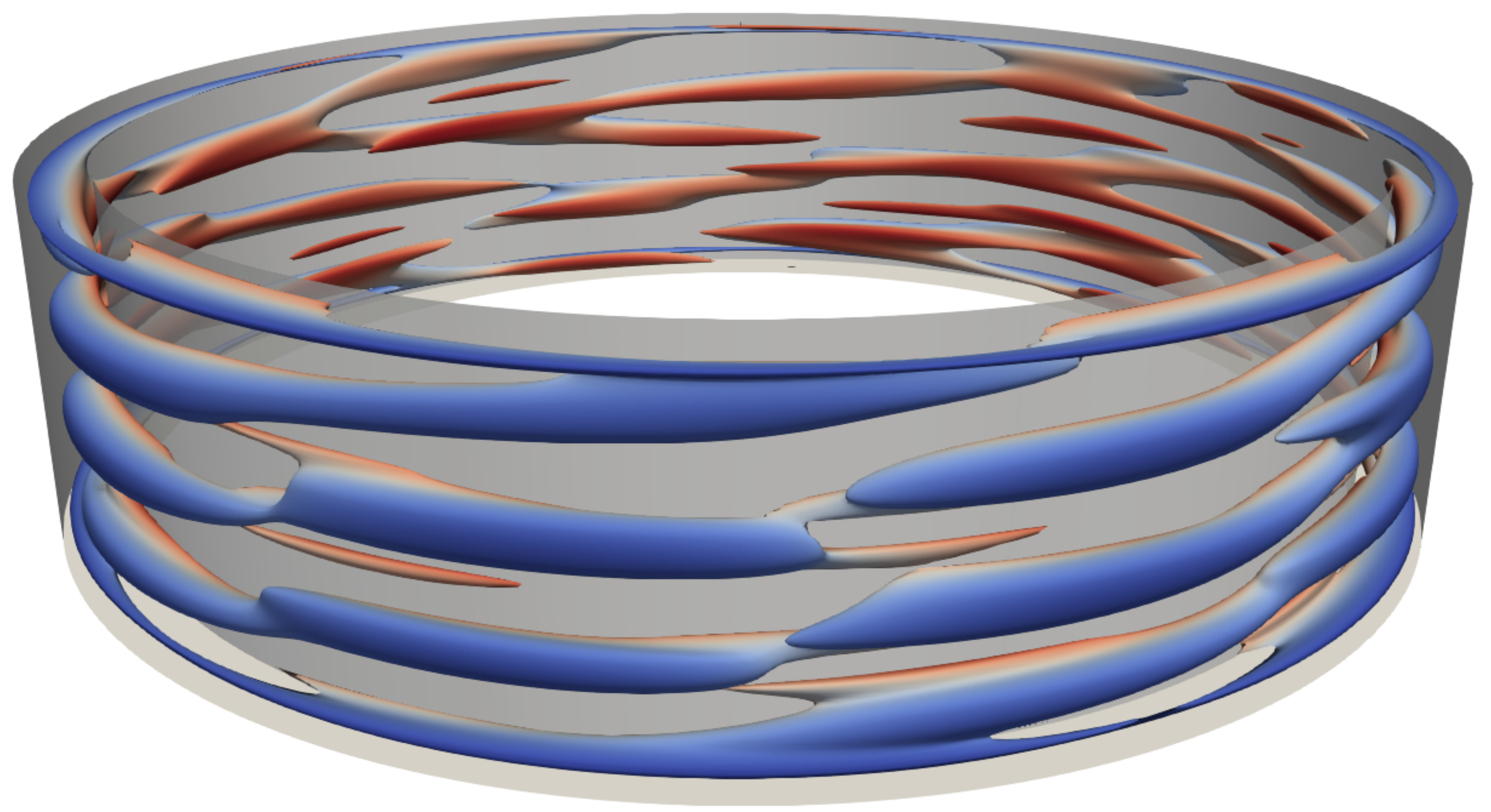}}
	\subfloat[]{\includegraphics[scale=0.139]{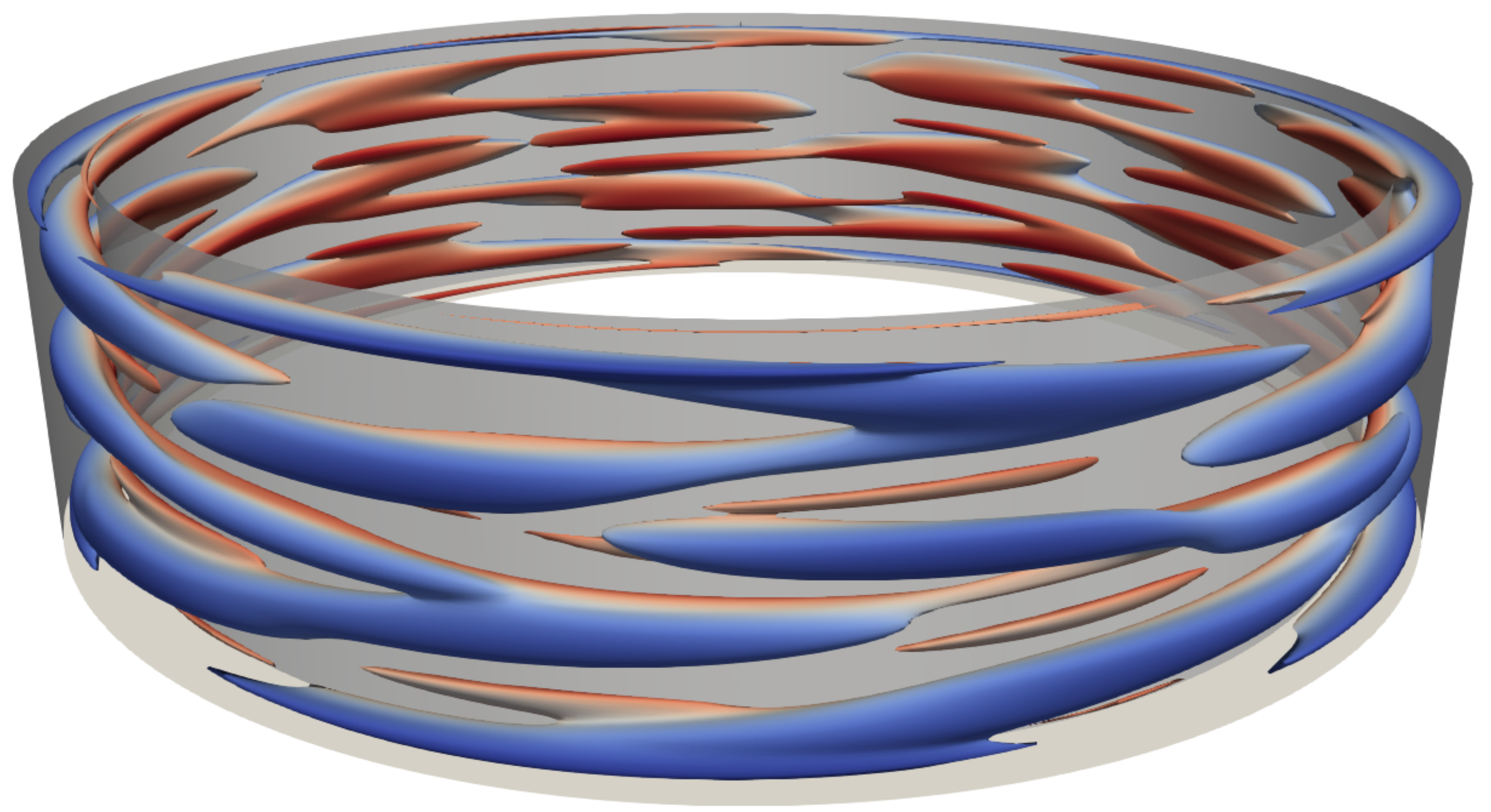}}
	\subfloat[]{\includegraphics[scale=0.139]{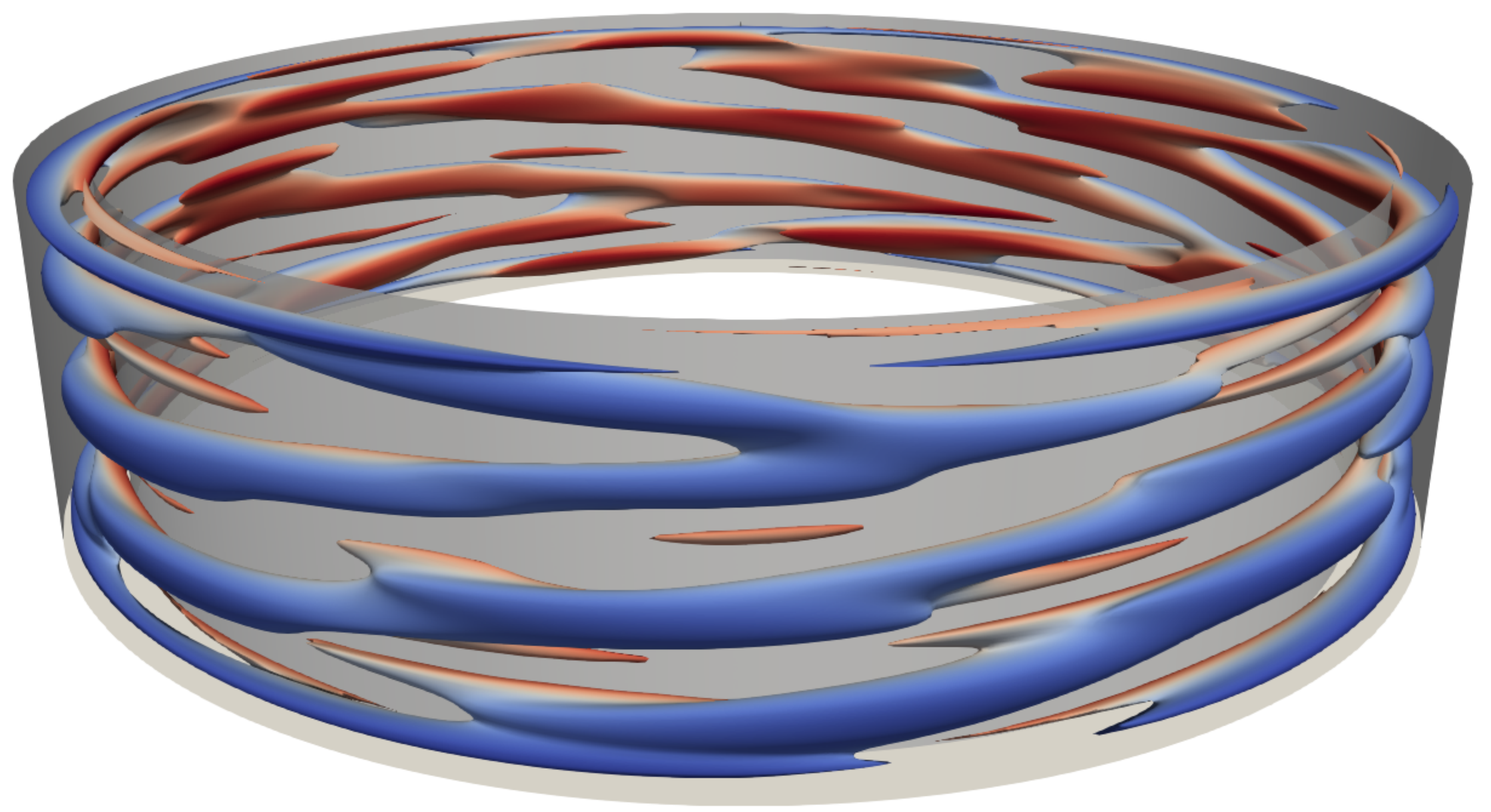}}
	\caption{\label{multiple_ips_flows}  Interpenetrating spirals at $Re_o =-1000$ and (a) $Re_i =620$, (b) $Re_i =625$, (c) $Re_i =630$. Each panel shows isosurfaces of the radial velocity component; the color indicates the corresponding azimuthal velocity component. Red (blue) indicates flow in the same direction as the inner (outer) cylinder rotation. }
\end{figure}

To quantify how the critical disturbance amplitude depends on \Rei, we fixed the spatial profile of the disturbance by choosing the initial condition in the form of a homotopy
\begin{equation}
\label{eq:homotopy}
\mathbf{v} = (1-\alpha)\mathbf{v}^\textrm{lam} + \alpha\mathbf{v}^\textrm{IPS},
\end{equation}
where $\mathbf{v}^{\textrm{lam}}$ is the laminar flow at the given \Rei\ and $\mathbf{v}^{\textrm{IPS}}$ is a snapshot of the (nonaxisymmetric) IPS at $Re_i = 630$. As Figure \ref{multiple_ips_flows} illustrates, the structure of the IPS is fairly similar at different \Rei; thus, for the purpose of determining critical disturbance amplitudes, we considered it to be sufficient to compute $\mathbf{v}^{\textrm{IPS}}$ at a fixed \Rei. The homotopy parameter $0\le \alpha\le 1$ characterizes the magnitude of the disturbance; increasing $\alpha$ increases the disturbance amplitude. This particular choice of homotopy guarantees that initial conditions are divergence-free for any value of $\alpha$. 
 
\begin{figure}
	\centering
	\includegraphics[width=\textwidth]{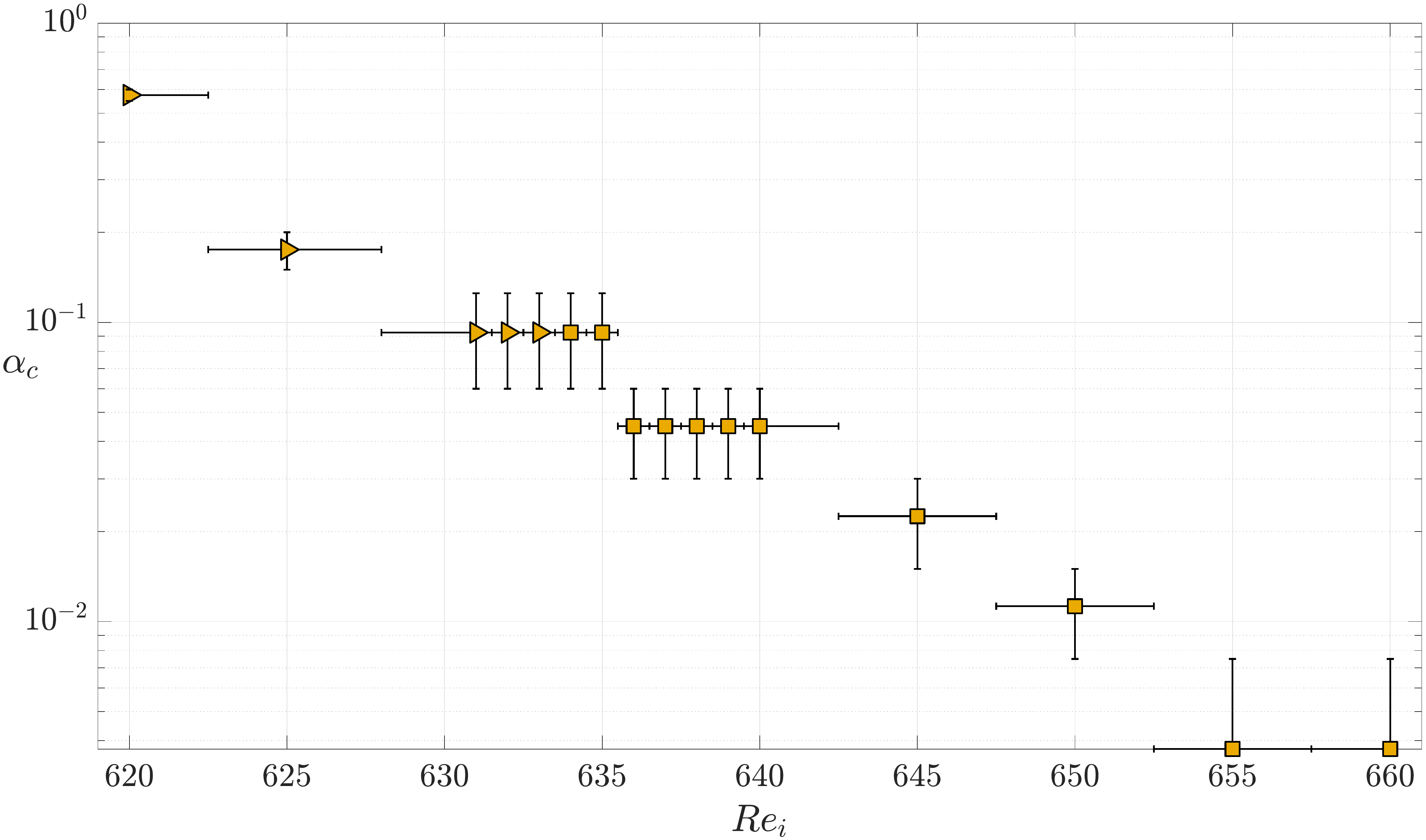}	
	\caption{\label{fig:alphatrans} The critical magnitude $\alpha_c$ of the homotopy parameter below which the flow relaminarizes at different \Rei. The symbols \protect\YellowTriangle{} and  \protect\YellowSquare{} denote that the flow at this \Rei\ transitions to IPS and turbulence, respectively, for $\alpha>\alpha_c$.
}
\end{figure}

For each \Rei\ we considered, a series of numerical simulations were performed with each simulation at a different value of $\alpha$; the simulations were run until the flow approached an asymptotic state. We then used bisection to determine the largest value of $\alpha$ at which the flow relaminarized. The critical value $\alpha_{c}$ is then defined as the midpoint between the two $\alpha$ values found that produce relaminarization and transition. The results are summarized in Figure \ref{fig:alphatrans} and suggest that the bifurcation at \Reic\ leading to the loss of stability of the laminar flow is subcritical, with $\alpha_c$ decreasing with increasing \Rei\ and vanishing at \Reic. Furthermore, for $Re_i\ge 634$, disturbances with $\alpha>\alpha_c$ lead to a transition to turbulence with IPS serving as a transient intermediate state. For $Re_i\le 633$, on the other hand,  disturbances with $\alpha>\alpha_c$ lead to a transition to stable IPS. We note that the value $Re_i=643\pm 2$ at which the transition to turbulence is found in experiment corresponds to $\alpha_c\approx 0.02$, suggesting the ambient disturbance level present in the experiments is fairly low.

\begin{figure}
	\centering
	\includegraphics[width=\columnwidth]{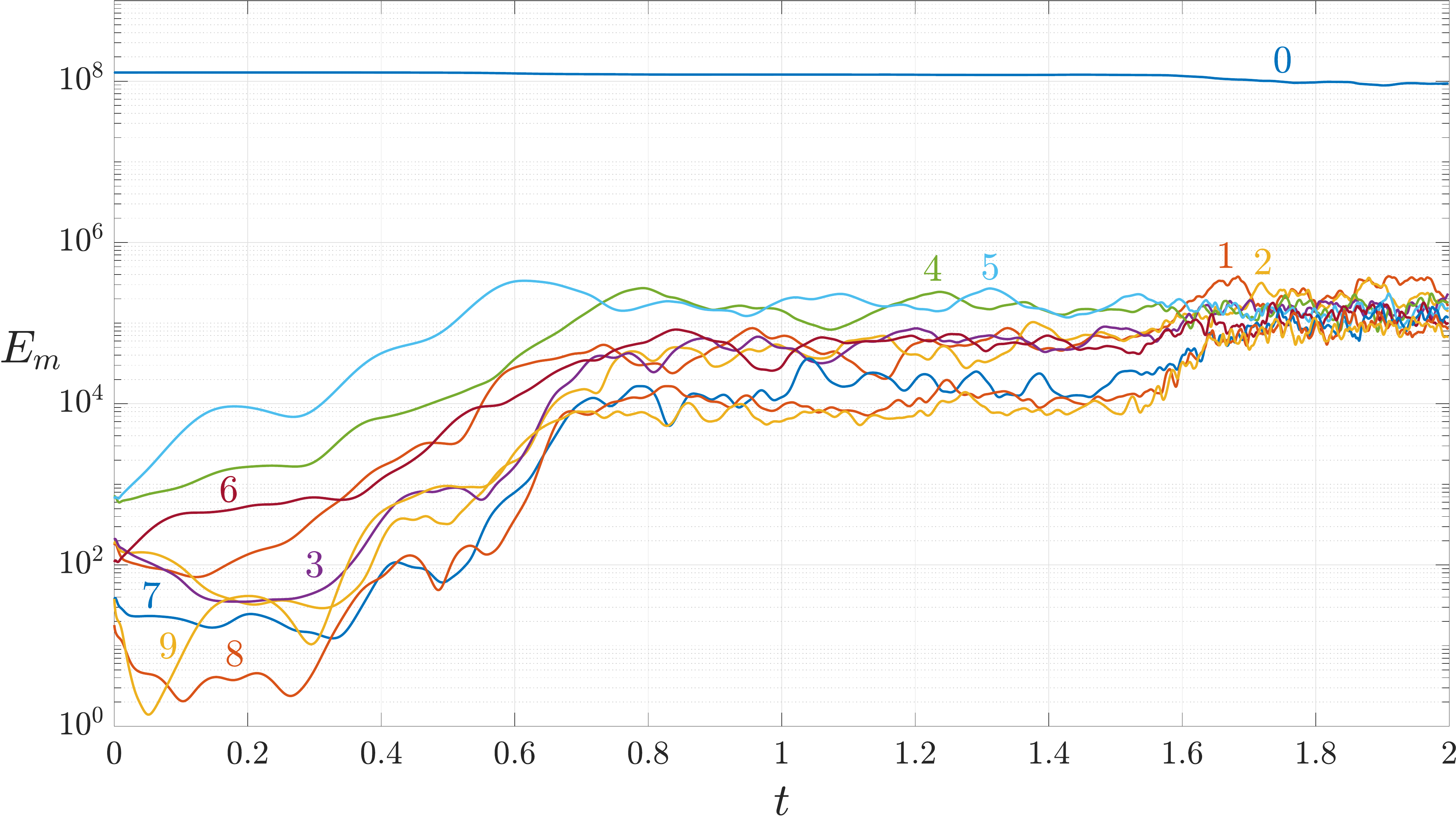}
	\caption{\label{kinen} The energy of the flow contained in the ten leading azimuthal Fourier modes during the transition from laminar flow to turbulence at $Re_i=640$. The mode numbers are shown next to each curve.}
\end{figure}

To illustrate the transition from laminar flow to turbulence, Figure \ref{kinen} shows the energies of the ten leading azimuthal Fourier modes ($m=0,\cdots,9$)
\begin{align}
E_m(t)=\int_{\eta/(1-\eta)}^{1/(1-\eta)} r\;dr \int_{0}^{\Gamma} |{\bf v}_m(t)|^2 \; dz,
\end{align}
where ${\bf v}_m$ is the $m$-Fourier component of the velocity field, for a representative numerical simulation at $Re_{i}=640$, where the initial condition was constructed using the homotopy \eqref{eq:homotopy} with $\alpha=0.06$. One can clearly see three distinct regimes: for $0<t\lessapprox0.7$ the perturbation about the laminar flow grows.  For $0.7\lessapprox t\lessapprox1.6$, the flow temporarily saturates into IPS where the mode energies remain roughly constant, with modes $m=4$ and $m=5$ dominating. Finally, for $t \gtrapprox 1.6 $, IPS give way to turbulence. The flows corresponding to the three regimes are qualitatively similar to those shown in Figure  \ref{flow_fields_transition}.

Numerical simulations find the same sequence of transitions as laboratory experiments when the flow is initially turbulent.  Decreasing \Rei\ first leads to a transition to stable IPS at  $Re_i =  623.5\pm 0.5$. From  stable IPS, increasing \Rei\ leads to a transition back to turbulence at $Re_i = 630.5 \pm 0.5$, while decreasing \Rei\ leads to a transition to the time-independent laminar state at $Re_i = 617.5\pm 0.5$. These numerically determined transition Reynolds numbers between IPS and turbulence and from IPS to laminar are quantitatively in agreement with those found in laboratory experiments, as illustrated in Table \ref{values}. Due to the subcritical nature of the transition between laminar and turbulence, however, the appropriate choice of the finite amplitude perturbation, $\alpha$, is required. 

The protocol for determining \Rei\ for transition from turbulence to IPS is as follows: 
We started with verifying that turbulence persists at $Re_i =640$ by evolving the flow for a time interval $5.264\,\tau$. 
Then we ramped down \Rei\ in increments of $\Delta \Rei=5$ and evolved the flow for the same interval to determine whether a transition occurred.
Once a transition was detected (at $Re_i =620$), we re-initialized the flow using the final state of the simulation at $Re_i =625$,  decreased the Reynolds number by $\Delta Re_{i}=1$, and evolved the flow for a further $5.264\,\tau$. The procedure was repeated with $\Delta Re_{i}=2,3,\cdots$ until a transition was found.

A similar protocol was used for the two transitions from stable IPS. In these cases, we verified that stable IPS persists at $Re_i =620$ and 630. The final states of the simulation at $Re_i =630$ (or $Re_i =620$) were evolved for $5.264\,\tau$ at a fixed \Rei\ that was increased (or decreased) by $\Delta \Rei=1,2,3,\cdots$ until transition to turbulence (or laminar flow) was found.  Note that, in all of these cases, only one simulation was performed and the finest resolution was $\Delta \Rei=1$, which determines the accuracy of the values reported in Table \ref{values}.

Given ample experimental evidence that the transitions between turbulence and IPS are probabilistic, we did not investigate these transitions numerically in more detail. For the transition from IPS to laminar flow, however, experiments did not conclusively determine the nature of the transition. We therefore performed an additional numerical investigation of this transition by evolving IPS at a number of fixed \Rei\ in the range $(617,618)$. While most of the results were consistent with a transition threshold found previously, there were a few outliers. 
In particular, we found that evolving IPS for $5.264\,\tau$ at $Re_i =617.8125$ does not result in a transition to a laminar flow, although eventually the flow does relaminarize. 
This result shows that the transition from IPS to laminar flow also appears to have a probabilistic nature and does not correspond to a bifurcation which would have resulted in a sharp transition boundary.

\begin{figure}
	\centering
	\subfloat[]{\includegraphics[trim={0cm 5cm -1cm 5cm},clip,height=4cm]{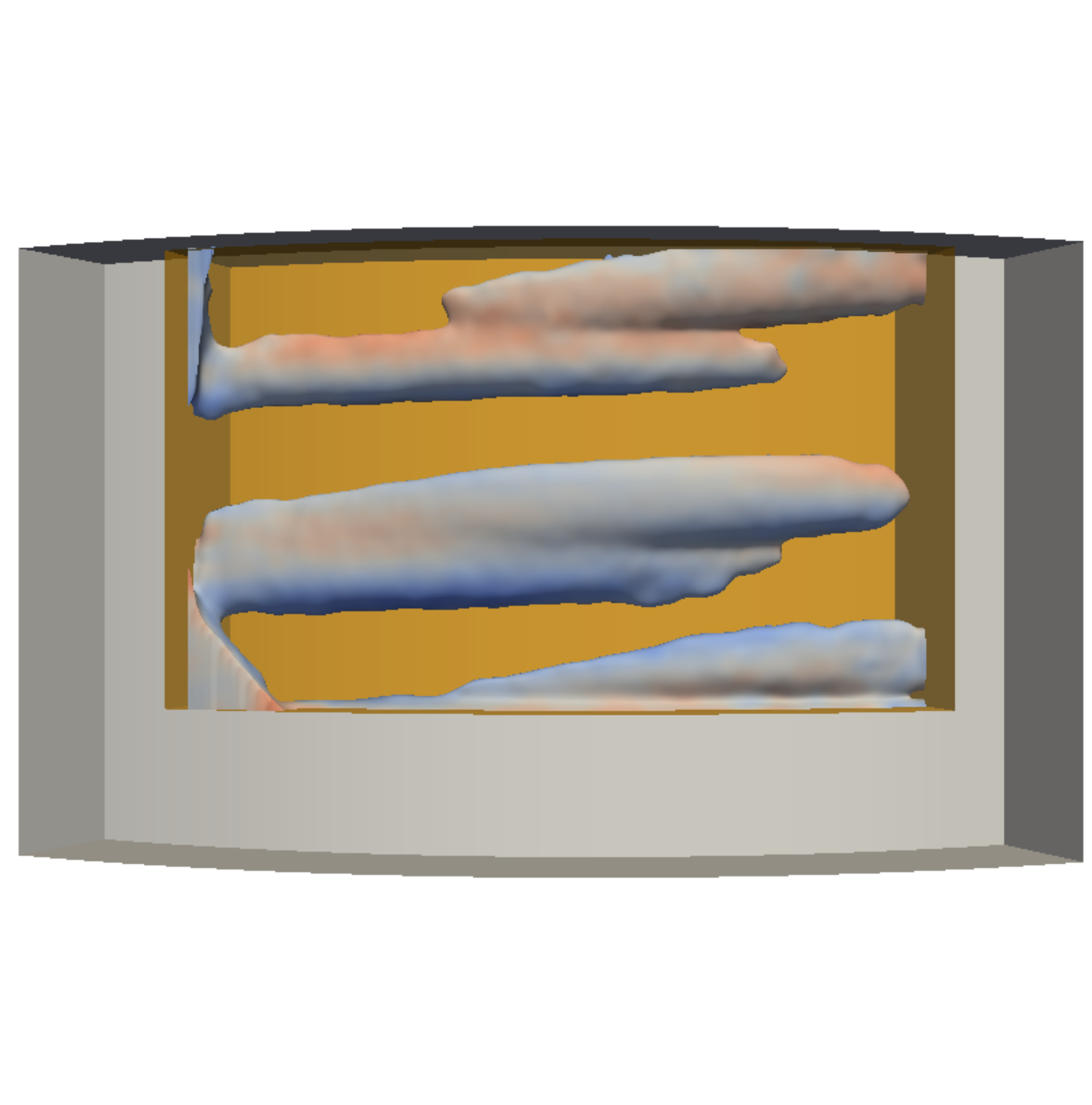}}
	\subfloat[]{\includegraphics[trim={-1cm 5cm 0cm 5cm},clip,height=4cm]{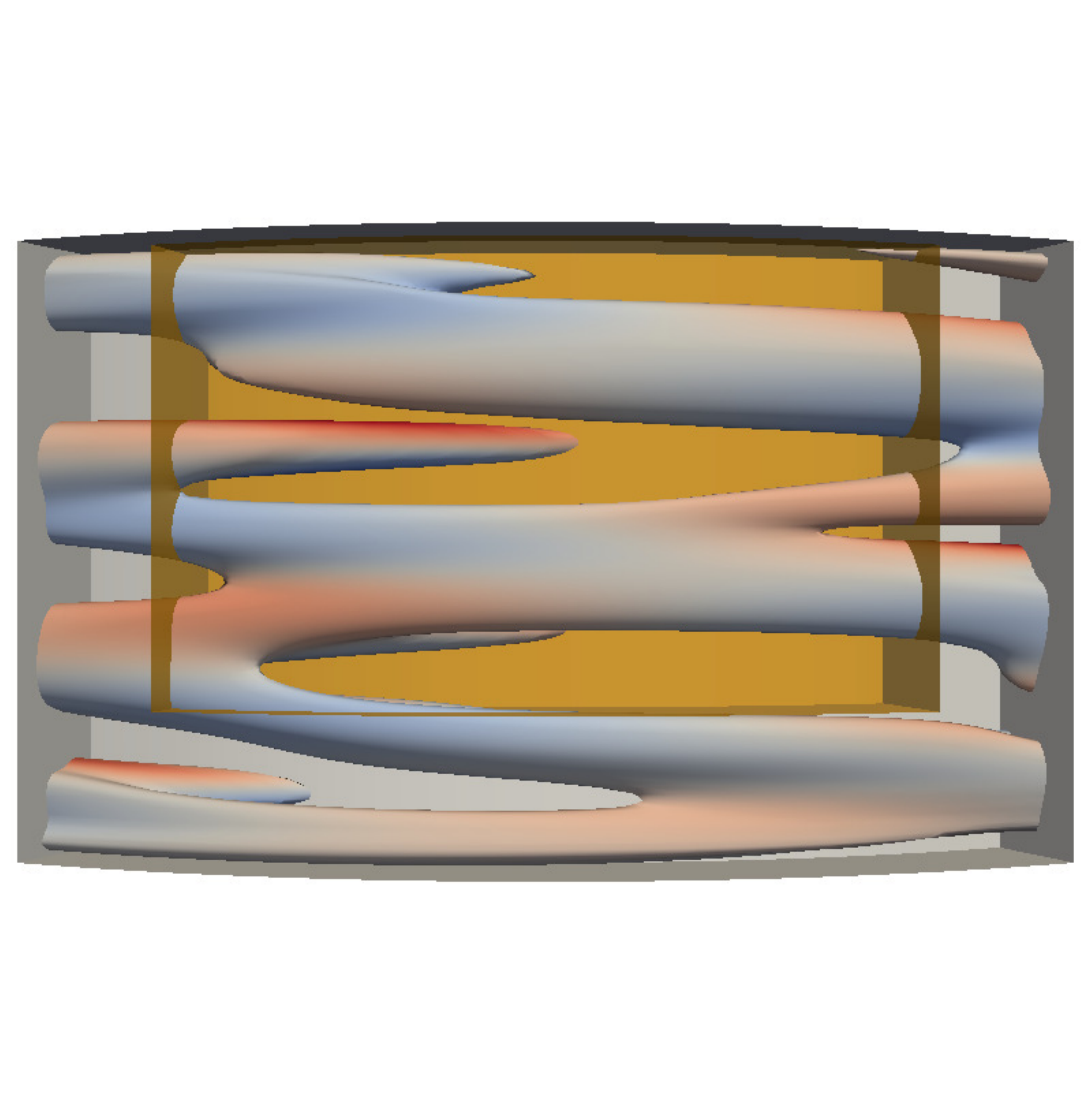}}	
		\caption{\label{flow_fields_IPS}  A snapshot of interpenetrating spirals in (a) a tomo PIV experiment and  (b) DNS. Each image shows a single isosurface of the perturbation field, $\widetilde{\mathbf{v}}_\theta$, for $Re_i =625$ and $Re_o =-1000$ inside a cylindrical subvolume. The color indicates the corresponding azimuthal velocity component. Red (blue) indicates flow in the same direction as the inner (outer) cylinder rotation. The shaded orange rectangular box represents the region probed by tomo PIV, which spans approximately 10\;\% of the flow domain volume.
}
\end{figure}

\begin{figure}
	\centering
	\subfloat[]{\includegraphics[trim={0cm 5cm -1cm 5cm},clip,height=4cm]{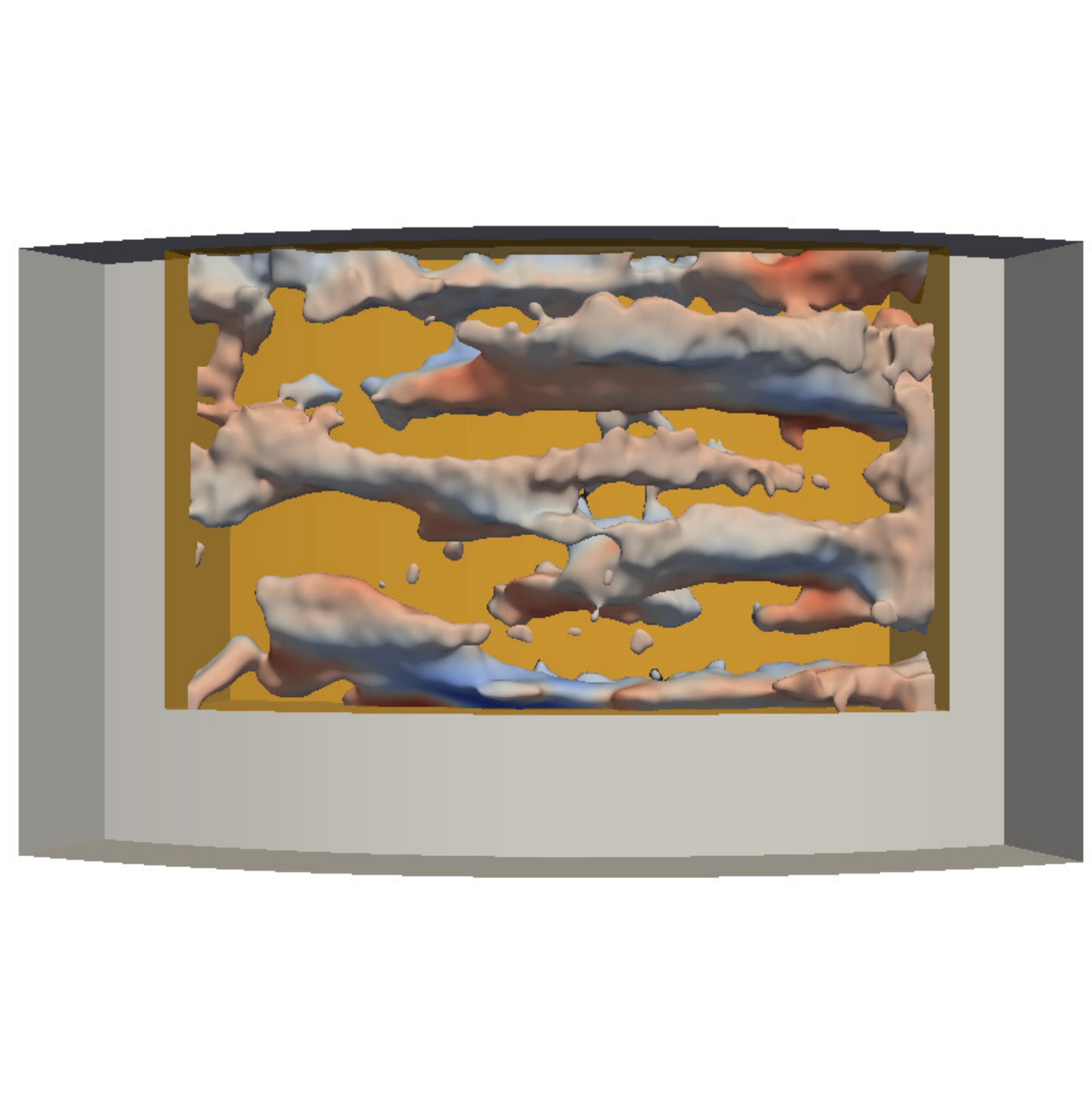}}
	\subfloat[]{\includegraphics[trim={-1cm 5cm 0cm 5cm},clip,height=4cm]{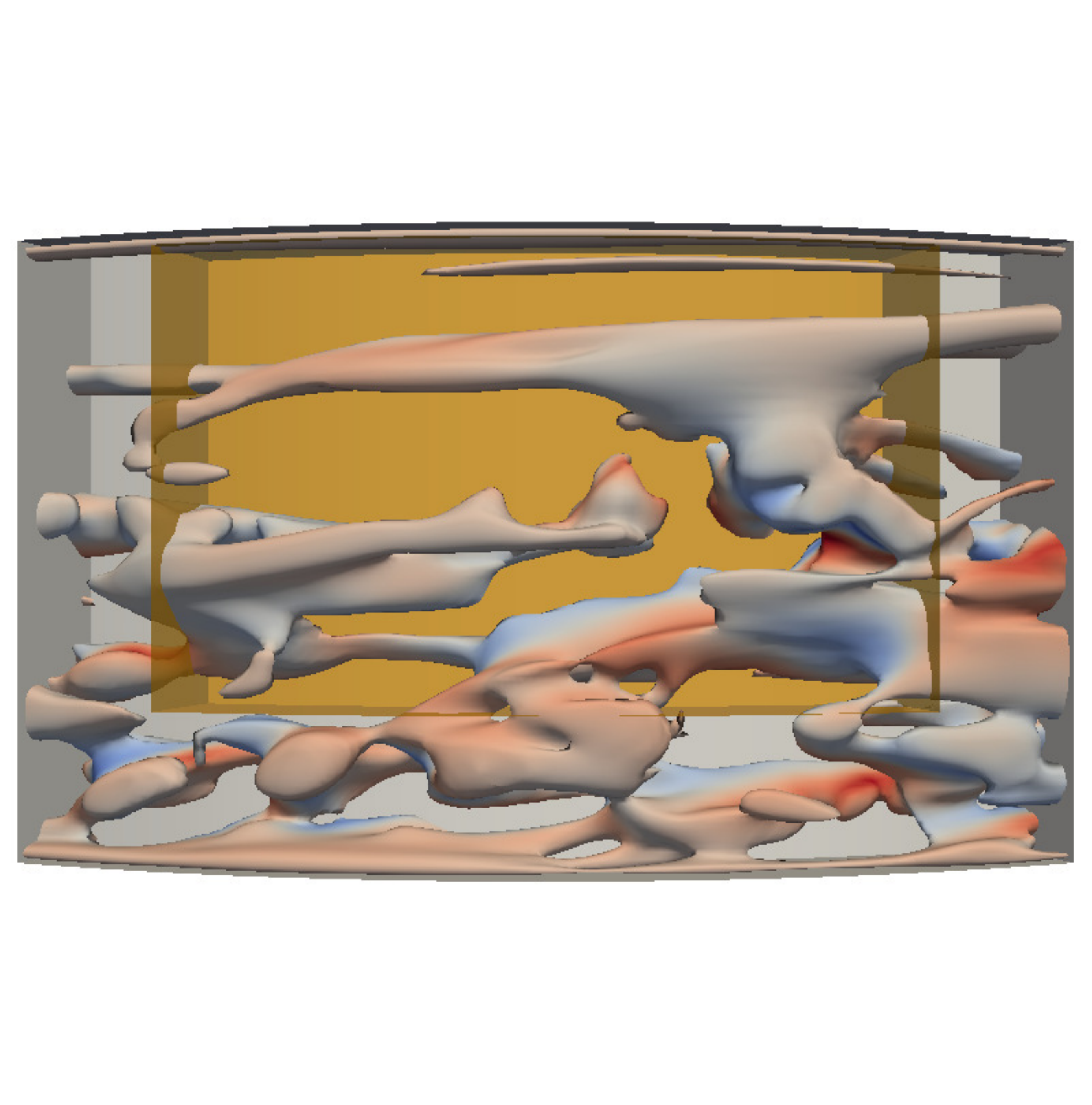}}	
	\caption{\label{flow_fields_Turb} A snapshot of a turbulent flow in experiment (a) and DNS (b). Each image shows a single isosurface of the perturbation field, $\widetilde{\mathbf{v}}_\theta$,  for $Re_i =650$ and $Re_o =-1000$ inside a cylindrical subvolume. The color indicates the corresponding azimuthal velocity component. Red (blue) indicates flow in the same direction as the inner (outer) cylinder rotation. The shaded orange rectangular box represents the region probed by tomo PIV, which spans approximately 10\;\% of the flow domain volume.
}
\end{figure}

\subsubsection{Flow Field Characterization}

Flow fields computed numerically also compare well with measurements from laboratory experiments. The stable IPS found in simulations and experiments exhibit a similar spatial structure (Figure \ref{flow_fields_IPS}).  Moreover, both experiments and simulations show that just above the onset of turbulence, the flow features localized patches of turbulence that co-exist with disordered spiral structures (cf. Figure \ref{flow_fields_transition}(c) and Figure \ref{flow_fields_Turb}). To quantitatively compare the flows in experiment and numerics, we computed the average energy $E$ corresponding to the $\theta$ component of the velocity perturbation $\widetilde{\mathbf{v}}={\bf v}-{\bf v}^\mathrm{lam}$ over a time interval $T=O(\tau)$ and region $\Omega$ in the $r,z$ plane at a fixed azimuthal location where experimental velocity measurements were available. Only the $\theta$-component of velocity was analyzed because $v_r$ and $v_z$ had increased noise due to the frame rates used in the PIV. The region $\Omega$ is bounded by the coordinates $r\in [\eta/(1-\eta),1/(1-\eta)]$  and $z/\Gamma\in [0.254,0.973]$, where $z$ is measured from the bottom of the flow domain. For the stable states (IPS and turbulence), the average energy was defined according to
\begin{equation}
\label{EQ:OrderParam}
    E = \frac{1}{T A}\int_0^Tdt \int_{\Omega} \: \widetilde{v}_\theta^2(t) \:drdz,
\end{equation}
where $A$ is the area of the cross section of $\Omega$.

\begin{figure}
	\centering
	\def\svgwidth{0.75\textwidth}
    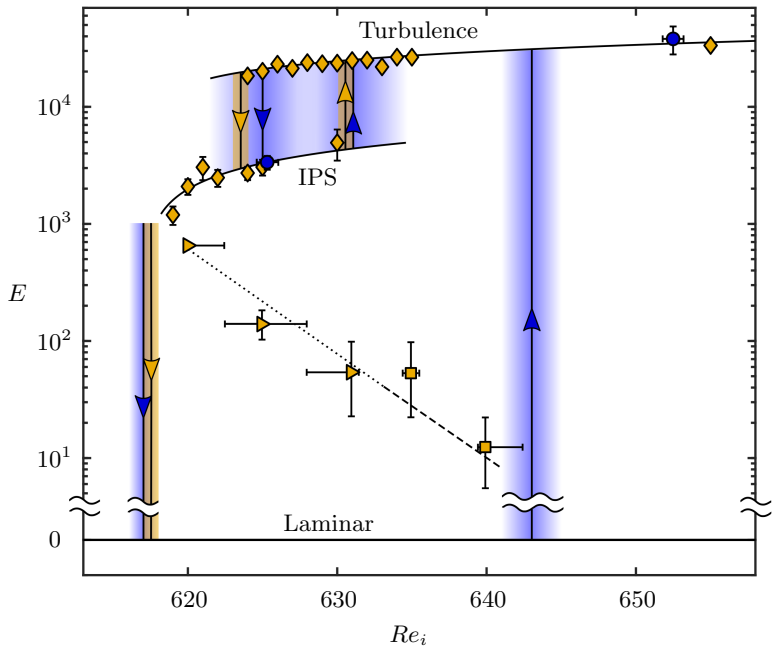
	\caption{\label{fig:bifur}  
	Transition map for the three flow states: laminar, IPS, and turbulence. Numerically computed values of $E$ for stable IPS and turbulence at various \Rei\ are plotted as \protect\YellowDiamond{} while the values calculated from experimental measurements are plotted as \protect\BlueCircle{}.  The gold and blue arrows indicate the values of \Rei\ at which transitions occur in simulation and experiment respectively. The shading around the arrows corresponds to the uncertainty in the transitional \Rei\ value. The \protect\YellowTriangle{} and \protect\YellowSquare{} represent the $E$ value associated with the critical magnitude of the finite amplitude perturbation, which corresponds to $\alpha=\alpha_c$, necessary to initiate transition from laminar flow to IPS and turbulence, respectively. All curves are to guide the eye only.
	}
\end{figure} 

With the use of $E$ as an order parameter, IPS and turbulence can be easily distinguished. A transition map  shown in Figure \ref{fig:bifur} quantifies both IPS and turbulence as well as the boundary of the basin of attraction of the laminar flow. The gold and blue arrows connecting the three flow states represent the value of \Rei\ where the flow transitions in numerics and experiment, respectively. The shading around these arrows represent the uncertainty in those values. For experiments, the uncertainty represents the repeatability of the transition while in the numerics the uncertainty represents the resolution of the steps in \Rei\ investigated. The diamonds and circles correspond to the value of $E$ computed from the numerics and experiment respectively. The triangles and squares correspond to the critical initial disturbance (for transition from laminar flow to IPS and turbulence, respectively) shown in Figure \ref{fig:alphatrans} in terms of the homotopy parameter $\alpha$. The average energy $E$ corresponding to the critical disturbance was computed by averaging over the azimuthal variable rather than time in \eqref{EQ:OrderParam} where $\widetilde{\bf v}=\alpha ({\bf v}^\mathrm{IPS}-{\bf v}^\mathrm{lam})$ according to \eqref{eq:homotopy} and $\alpha=\alpha_{c}$.

\section{Discussion} 
\label{sec:discussion}

Two distinct instabilities at play in counter-rotating TCF form the basis for a qualitative physical picture of turbulent transition.   In the limiting case where \Reo\ approaches zero (the outer cylinder is at rest), the laminar flow is subject to centrifugal instability when \Rei\ is sufficiently large.  By contrast, in the limiting case where \Rei\ approaches zero (the inner cylinder is at rest), the flow is centrifugally stable for all values of \Reo, but is subject to shear instability for \Reo\ sufficiently large.   Under counter-rotation, both instability mechanisms can, roughly speaking, be thought of as operative in distinct spatial regions, separated by a ``nodal surface'' where the azimuthal velocity component is zero.  On the side of the nodal surface nearer the inner cylinder, the azimuthal velocity component is decreasing with increasing radial distance from the inner cylinder, thereby, providing a necessary condition for centrifugal instability in this (inner) flow region.  On the side of the nodal surface nearer the outer cylinder, centrifugal instability is ruled out since the azimuthal velocity component is increasing with increasing radial distance from the inner cylinder; however, shear flow instabilities remain as a possibility in this (outer) flow region.

Prior work in large-aspect-ratio counter-rotating TCF has described a scenario in which the interplay between the inner and outer flow regions leads to turbulent transition.   When \Reo\ is fixed and sufficiently large in magnitude and \Rei\ is increased quasi-statically, the primary instability of the laminar flow leads to the formation of stable spiral flows \citep{Coles1967, Andereck1986, Eckhardt1995,Goharzadeh2001}, which are mainly confined to the centrifugally-unstable inner region and qualitatively similar to IPS described in the present paper.  Simulations with periodic axial boundary conditions \citep{Coughlin1996} showed that, as \Rei\ is increased beyond the primary instability, the spiral flow in the inner region increasingly disturbs the centrifugally-stable outer region.  Coughlin \& Marcus found that, beyond a certain \Rei, the disturbance amplitude becomes strong enough to trigger a shear instability in the outer layer leading to turbulence.   This transition scenario is in qualitative agreement with experimental observation in TCF with moderate-to-large aspect ratios ($17 \le \Gamma \le 46$) \citep{Hamill1995}.

\begin{figure}
	\centering
	\subfloat[]{\includegraphics[height=0.49\columnwidth,angle=90]{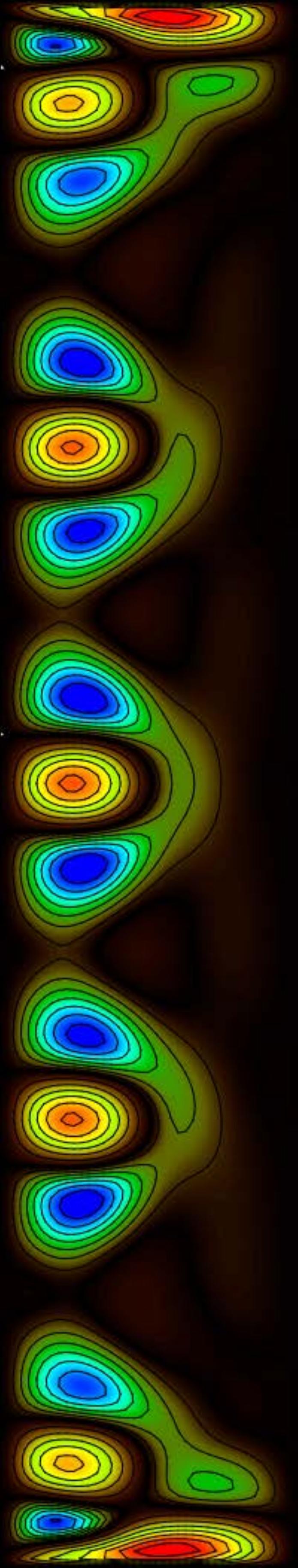}}\hspace{1mm}
	\subfloat[]{\includegraphics[height=0.49\columnwidth,angle=90]{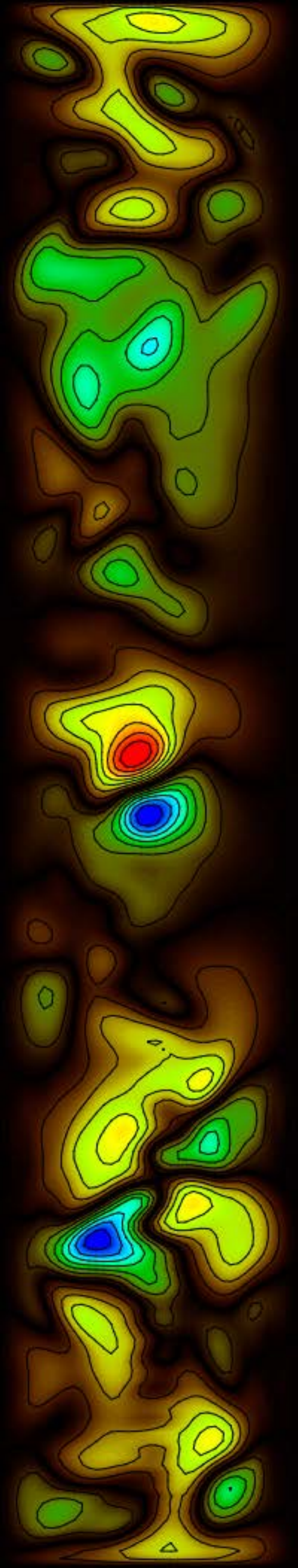}}
	\caption{\label{moneyplane} Typical snapshots of the radial velocity in the constant $\theta$ plane for (a) IPS and (b) turbulence in the numerical simulation at $Re_i=637$ and $Re_o=-1000$. The $r$ direction is vertical and $z$ is horizontal, with the inner cylinder at the bottom. Red (blue) indicates positive (negative) values of $v_r$.
}
\end{figure}

Our experimental results suggest the interactions between inner and outer flow regions also play a central role in transition in small-aspect-ratio TCF, with the important difference that transition from laminar flow leads directly to turbulence facilitated by the temporary appearance of IPS.  The laminar state with \Reo\ fixed exhibits a subcritical rather than a supercritical instability as \Rei\ increased quasi-statically. Consequently, as \Rei\ increases, the laminar flow undergoes a finite amplitude instability leading to growth of a spiral flow mostly confined to the inner region (cf. Figure \ref{moneyplane}(a)). However, unlike the large-aspect-ratio case, the emerging spiral states are transient, with the flow in the inner region generating disturbances of sufficient amplitude to trigger shear instability in the outer region leading to turbulence (cf. Figure \ref{moneyplane}(b)). Stable IPS do exist in our system at lower \Rei, but are disconnected from the axisymmetric laminar solution. The transition from stable IPS to turbulence appears similar to the large-aspect-ratio case.

This transition, as well as the reverse transition from turbulence to stable IPS are similar from a dynamical systems standpoint.  The results shown in Figure \ref{fig:stats}(a) for the decay from turbulence is reminiscent of earlier observations of transitions from turbulence in pipe flows \citep{Faisst2004, Peixinho2006, Avila2011} and in large-aspect-ratio TCF driven solely by outer cylinder rotation \citep{Borrero2010}.  In all these cases, the exponential decay from turbulence is suggestive of a memoryless process in which, from a state space viewpoint, the transient character of turbulence is captured by a finite-time escape from a chaotic repeller to a qualitatively different solution. Interestingly, in all previous work, once the turbulent transients had disappeared, the flow relaminarized; by contrast, our results demonstrate, for the first time, that turbulence gives way to another chaotic state (IPS).  Moreover, Figure \ref{fig:stats}(b) suggests the transition from IPS to turbulence exhibits a similar character, so that for sufficiently large \Rei, IPS are described by a chaotic repeller and the flow transitions to a different chaotic state (turbulence). In this regard, the origin of the IPS-to-turbulent transition observed here is quite different from the linear secondary instability mechanism proposed earlier \citep{Coughlin1996}. The change of the nature of the chaotic set underlying IPS from an attractor at lower \Rei\ to a repeller at higher \Rei\ also explains the transient appearance of IPS during the transition from laminar to turbulent flow. 

Prior work has demonstrated the chaotic behavior of IPS arises from competition between spiral modes of different wavenumbers and helicities \citep{Andereck1986, Hamill1995, Coughlin1996}.  Our observations of IPS were made in a TCF apparatus with an aspect ratio substantially smaller than that employed in earlier studies of IPS; thus, axial confinement effects in our work hinders clear identification of distinct spiral modes at play in IPS.  Nevertheless, we speculate that chaos in IPS observed here originates from qualitatively similar mode interactions as that found in larger aspect ratio studies.

\section{Conclusion} 
\label{sec:conclusion}

Our experimental and numerical results indicate that, for suitable parameter values, Taylor-Couette flow can exhibit some key characteristics commonly observed in the transition to turbulence in other shear flows.   The transition from laminar flow to turbulence is subcritical, like that observed for flows in channels and boundary layers.  In particular, when transition is probed by increasing \Rei\ sufficiently slowly (see Sec. \ref{sec:results_lam_turb}), structured, non-turbulent flows (transient interpenetrating sprials) mediate the transition to turbulence in Taylor-Couette flow in a manner that is analogous to the role of Tollmein-Schlicting (TS) waves in the transition to turbulence in channels and boundary layers in low-noise environments.  Moreover, when transition is probed by increasing \Rei\ sufficiently rapidly with relatively large step changes $\Delta\Rei$, the corresponding disturbances to the laminar flow cause turbulent transition at values of  \Rei\ that decrease with increasing $\Delta\Rei$.  Under these conditions, the dependence of the laminar-turbulent transition on disturbance amplitude is reminiscent of bypass transition scenarios observed in channels and boundary layers when the ambient (free-stream) turbulence intensity is sufficiently large.

However, there are also significant differences between Taylor-Couette flow and other canonical shear flows: the physical instability mechanism of IPS (centrifugal instability) differs from the mechanisms for TS waves;  TCF transition does not feature a linear growth regime like that found for TS waves in channel and boundary layer transition; stable, nonlinearly-saturated TS waves are never observed in channels/boundary layers, unlike the stable IPS observed in TCF.

While TCF is a closed flow while, most commonly, the subcritical laminar-turbulent transition is studied in open flows. 
Nevertheless, the highly reproducible character of the transition, with close correspondence between numerics and experiment in TCF offer the opportunity to explore in great detail shear flow transition behaviors that may show up generally in a variety of settings.
One such opportunity for future study emerging from recent theoretical and experimental work suggests that the dynamics of turbulent flows are guided by particular unstable solutions to the Navier-Stokes equation \citep{Hof2004,suri2017}.   This work suggests that selected solutions with simple temporal behavior (e.g., equilibria, limit cycle oscillations) exhibit spatial structures that are strikingly similar to well-known patterns (coherent structures) that have long been known to play a central role in turbulence; moreover, a suitable selection of such solutions (known as \emph{exact coherent structures}) can be harnessed to capture turbulent flow dynamics and statistics (e.g., average turbulent flow properties).   Our results suggest that exact coherent structures with spiral spatial structures could play a role in mediating laminar-turbulent transition in counter-rotating Taylor-Couette flow.

\section{Acknowledgements} 
\label{sec:acknowledgements}
The authors would like to acknowledge the financial support by the Army Research Office under grants No. W911NF-15-10471 and  W911NF-16-10281. D.B.-E. gratefully acknowledges the support of the M.J.Murdock Charitable Trust (Award \# 2015214) and the Kresge Endowment at Willamette University. We are grateful to Marc Avila for sharing the Taylor-Couette DNS solver used here with us and for many useful discussions.

\clearpage
\bibliographystyle{jfm}
\bibliography{TR_Transition}

\end{document}

%% file: Exp_Transition_Full.pdf_tex
\begingroup%
  \makeatletter%
  \providecommand\color[2][]{%
    \errmessage{(Inkscape) Color is used for the text in Inkscape, but the package 'color.sty' is not loaded}%
    \renewcommand\color[2][]{}%
  }%
  \providecommand\transparent[1]{%
    \errmessage{(Inkscape) Transparency is used (non-zero) for the text in Inkscape, but the package 'transparent.sty' is not loaded}%
    \renewcommand\transparent[1]{}%
  }%
  \providecommand\rotatebox[2]{#2}%
  \ifx\svgwidth\undefined%
    \setlength{\unitlength}{716.21860352bp}%
    \ifx\svgscale\undefined%
      \relax%
    \else%
      \setlength{\unitlength}{\unitlength * \real{\svgscale}}%
    \fi%
  \else%
    \setlength{\unitlength}{\svgwidth}%
  \fi%
  \global\let\svgwidth\undefined%
  \global\let\svgscale\undefined%
  \makeatother%
  \begin{picture}(1,0.7269882)%
    \put(0.06550129,0.09714712){\color[rgb]{0,0,0}\makebox(0,0)[lt]{\begin{minipage}{0.23849953\unitlength}\raggedright \large -4000\end{minipage}}}%
    \put(0.25365132,0.0958217){\color[rgb]{0,0,0}\makebox(0,0)[lt]{\begin{minipage}{0.23849953\unitlength}\raggedright \large -3000\end{minipage}}}%
    \put(0.44199769,0.09566271){\color[rgb]{0,0,0}\makebox(0,0)[lt]{\begin{minipage}{0.23849953\unitlength}\raggedright \large -2000\end{minipage}}}%
    \put(0.63042854,0.09518888){\color[rgb]{0,0,0}\makebox(0,0)[lt]{\begin{minipage}{0.23849953\unitlength}\raggedright \large -1000\end{minipage}}}%
    \put(0.86,0.10952561){\color[rgb]{0,0,0}\makebox(0,0)[lt]{\begin{minipage}{0.23849953\unitlength}\raggedright \large 0\end{minipage}}}%
    \put(0.86,0.21765295){\color[rgb]{0,0,0}\makebox(0,0)[lt]{\begin{minipage}{0.23849953\unitlength}\raggedright \large 500\end{minipage}}}%
    \put(0.86,0.42428897){\color[rgb]{0,0,0}\makebox(0,0)[lt]{\begin{minipage}{0.23849953\unitlength}\raggedright \large 1000\end{minipage}}}%
    \put(0.86,0.62721849){\color[rgb]{0,0,0}\makebox(0,0)[lt]{\begin{minipage}{0.23849953\unitlength}\raggedright \large 1500\end{minipage}}}%
    \put(0.45,0.06){\color[rgb]{0,0,0}\makebox(0,0)[lt]{\begin{minipage}{0.13621299\unitlength}\raggedright \Large $Re_{o}$\end{minipage}}}%
    \put(0.92845909,0.43){\color[rgb]{0,0,0}\makebox(0,0)[lt]{\begin{minipage}{0.13621299\unitlength}\raggedright \Large $Re_{i}$\end{minipage}}}%
    \put(0.48,0.18){\color[rgb]{0,0,0}\makebox(0,0)[lt]{\begin{minipage}{0.41327206\unitlength}\raggedright \large Linear stability\end{minipage}}}%
    \put(0.45,0.15){\color[rgb]{0,0,0}\makebox(0,0)[lt]{\begin{minipage}{0.41327206\unitlength}\raggedright \large boundary for $\Gamma=\infty$\end{minipage}}}%
    \put(0.13,0.4541771){\color[rgb]{0,0,0}\makebox(0,0)[lt]{\begin{minipage}{0.41327206\unitlength}\raggedright \large Turbulent to laminar\end{minipage}}}%
    \put(0.26151815,0.64691381){\color[rgb]{0,0,0}\makebox(0,0)[lt]{\begin{minipage}{0.41327206\unitlength}\raggedright \large Laminar to turbulent\end{minipage}}}%
    \put(0,0){\includegraphics[width=\unitlength,page=1]{Exp_Transition_Full}}%
  \end{picture}%
\endgroup%

%% file: BothPlots.pdf_tex
\begingroup%
  \makeatletter%
  \providecommand\color[2][]{%
    \errmessage{(Inkscape) Color is used for the text in Inkscape, but the package 'color.sty' is not loaded}%
    \renewcommand\color[2][]{}%
  }%
  \providecommand\transparent[1]{%
    \errmessage{(Inkscape) Transparency is used (non-zero) for the text in Inkscape, but the package 'transparent.sty' is not loaded}%
    \renewcommand\transparent[1]{}%
  }%
  \providecommand\rotatebox[2]{#2}%
  \newcommand*\fsize{\dimexpr\f@size pt\relax}%
  \newcommand*\lineheight[1]{\fontsize{\fsize}{#1\fsize}\selectfont}%
  \ifx\svgwidth\undefined%
    \setlength{\unitlength}{1165.19240011bp}%
    \ifx\svgscale\undefined%
      \relax%
    \else%
      \setlength{\unitlength}{\unitlength * \real{\svgscale}}%
    \fi%
  \else%
    \setlength{\unitlength}{\svgwidth}%
  \fi%
  \global\let\svgwidth\undefined%
  \global\let\svgscale\undefined%
  \makeatother%
  \begin{picture}(1,0.45314404)%
    \lineheight{1}%
    \setlength\tabcolsep{0pt}%
    \put(0.025,0){\includegraphics[width=\unitlength]{BothPlots.pdf}}%
    \put(0.0445,0.37275434){\color[rgb]{0,0,0}\makebox(0,0)[lt]{\lineheight{1.25}\smash{\begin{tabular}[t]{l}$10^0$\end{tabular}}}}%
    \put(0.0425,0.26168366){\color[rgb]{0,0,0}\makebox(0,0)[lt]{\lineheight{1.25}\smash{\begin{tabular}[t]{l}$10^{\shortminus 1}$\end{tabular}}}}%
    \put(0.0425,0.15060667){\color[rgb]{0,0,0}\makebox(0,0)[lt]{\lineheight{1.25}\smash{\begin{tabular}[t]{l}$10^{\shortminus 2}$\end{tabular}}}}%
    \put(0.0,0.23){\color[rgb]{0,0,0}\makebox(0,0)[lt]{\lineheight{1.25}\smash{\begin{tabular}[t]{l}$P$\end{tabular}}}}%
    \put(0.20,0.02529644){\color[rgb]{0,0,0}\makebox(0,0)[lt]{\lineheight{1.25}\smash{\begin{tabular}[t]{l}$2$\end{tabular}}}}%
    \put(0.355,0.02568618){\color[rgb]{0,0,0}\makebox(0,0)[lt]{\lineheight{1.25}\smash{\begin{tabular}[t]{l}$4$\end{tabular}}}}%
    \put(0.28,0.00045231){\color[rgb]{0,0,0}\makebox(0,0)[lt]{\lineheight{1.25}\smash{\begin{tabular}[t]{l}$t$\end{tabular}}}}%
    \put(0.5225,0.37275434){\color[rgb]{0,0,0}\makebox(0,0)[lt]{\lineheight{1.25}\smash{\begin{tabular}[t]{l}$10^0$\end{tabular}}}}%
    \put(0.525,0.26168366){\color[rgb]{0,0,0}\makebox(0,0)[lt]{\lineheight{1.25}\smash{\begin{tabular}[t]{l}$10^{\shortminus 1}$\end{tabular}}}}%
    \put(0.52,0.15260667){\color[rgb]{0,0,0}\makebox(0,0)[lt]{\lineheight{1.25}\smash{\begin{tabular}[t]{l}$10^{\shortminus 2}$\end{tabular}}}}%
    \put(0.71,0.02529644){\color[rgb]{0,0,0}\makebox(0,0)[lt]{\lineheight{1.25}\smash{\begin{tabular}[t]{l}$0.4$\end{tabular}}}}%
    \put(0.865,0.02568618){\color[rgb]{0,0,0}\makebox(0,0)[lt]{\lineheight{1.25}\smash{\begin{tabular}[t]{l}$0.8$\end{tabular}}}}%
    \put(0.755,0.00045231){\color[rgb]{0,0,0}\makebox(0,0)[lt]{\lineheight{1.25}\smash{\begin{tabular}[t]{l}$t$\end{tabular}}}}%
    \put(0.26,0.34556093){\color[rgb]{0,0,0}\makebox(0,0)[lt]{\lineheight{1.25}\smash{\begin{tabular}[t]{l}Turbulence to IPS\end{tabular}}}}%
    \put(0.74,0.34556093){\color[rgb]{0,0,0}\makebox(0,0)[lt]{\lineheight{1.25}\smash{\begin{tabular}[t]{l}IPS to Turbulence\end{tabular}}}}%
		
		\put(0.268,-0.035){\color[rgb]{0,0,0}\makebox(0,0)[lt]{\lineheight{1.25}\smash{\begin{tabular}[t]{l}(a)\end{tabular}}}}
	\put(0.743,-0.035){\color[rgb]{0,0,0}\makebox(0,0)[lt]{\lineheight{1.25}\smash{\begin{tabular}[t]{l}(b)\end{tabular}}}}
  \end{picture}%
	
	\vspace{0.25in}
	
\endgroup%

%% file: order_parameter_plot.pdf_tex
\begingroup%
  \makeatletter%
  \providecommand\color[2][]{%
    \errmessage{(Inkscape) Color is used for the text in Inkscape, but the package 'color.sty' is not loaded}%
    \renewcommand\color[2][]{}%
  }%
  \providecommand\transparent[1]{%
    \errmessage{(Inkscape) Transparency is used (non-zero) for the text in Inkscape, but the package 'transparent.sty' is not loaded}%
    \renewcommand\transparent[1]{}%
  }%
  \providecommand\rotatebox[2]{#2}%
  \newcommand*\fsize{\dimexpr\f@size pt\relax}%
  \newcommand*\lineheight[1]{\fontsize{\fsize}{#1\fsize}\selectfont}%
  \ifx\svgwidth\undefined%
    \setlength{\unitlength}{621.51274109bp}%
    \ifx\svgscale\undefined%
      \relax%
    \else%
      \setlength{\unitlength}{\unitlength * \real{\svgscale}}%
    \fi%
  \else%
    \setlength{\unitlength}{\svgwidth}%
  \fi%
  \global\let\svgwidth\undefined%
  \global\let\svgscale\undefined%
  \makeatother%
  \begin{picture}(1,0.86807433)%
    \lineheight{1}%
    \setlength\tabcolsep{0pt}%
    \put(0,0){\includegraphics[width=\unitlength]{order_parameter_plot.pdf}}%
    \put(0.38071189,0.60498471){\color[rgb]{0,0,0}\makebox(0,0)[lt]{\lineheight{0}\smash{\begin{tabular}[t]{l}IPS\end{tabular}}}}%
    \put(0.76227654,0.86807433){\color[rgb]{0,0,0}\makebox(0,0)[lt]{\lineheight{0}\smash{\begin{tabular}[t]{l} \end{tabular}}}}%
    \put(0.45937745,0.79571316){\color[rgb]{0,0,0}\makebox(0,0)[lt]{\lineheight{0}\smash{\begin{tabular}[t]{l}Turbulence\end{tabular}}}}%
    \put(0.36143367,0.15426302){\color[rgb]{0,0,0}\makebox(0,0)[lt]{\lineheight{0}\smash{\begin{tabular}[t]{l}Laminar\end{tabular}}}}%
    \put(0.00108417,0.45350279){\color[rgb]{0,0,0}\makebox(0,0)[lt]{\lineheight{0}\smash{\begin{tabular}[t]{l}$E$\end{tabular}}}}%
    \put(0.49994623,0.01002157){\color[rgb]{0,0,0}\makebox(0,0)[lt]{\lineheight{0}\smash{\begin{tabular}[t]{l}$Re_i$\end{tabular}}}}%
    \put(0.03914288,0.69560838){\color[rgb]{0,0,0}\makebox(0,0)[lt]{\lineheight{1.25}\smash{\begin{tabular}[t]{l}$10^4$\end{tabular}}}}%
    \put(0.03914288,0.54296071){\color[rgb]{0,0,0}\makebox(0,0)[lt]{\lineheight{1.25}\smash{\begin{tabular}[t]{l}$10^3$\end{tabular}}}}%
    \put(0.03914288,0.3903131){\color[rgb]{0,0,0}\makebox(0,0)[lt]{\lineheight{1.25}\smash{\begin{tabular}[t]{l}$10^2$\end{tabular}}}}%
    \put(0.03914288,0.23766553){\color[rgb]{0,0,0}\makebox(0,0)[lt]{\lineheight{1.25}\smash{\begin{tabular}[t]{l}$10^1$\end{tabular}}}}%
    \put(0.05498241,0.13346789){\color[rgb]{0,0,0}\makebox(0,0)[lt]{\lineheight{1.25}\smash{\begin{tabular}[t]{l}$0$\end{tabular}}}}%
    \put(0.21445216,0.0558586){\color[rgb]{0,0,0}\makebox(0,0)[lt]{\lineheight{1.25}\smash{\begin{tabular}[t]{l}$620$\end{tabular}}}}%
    \put(0.40967308,0.0558586){\color[rgb]{0,0,0}\makebox(0,0)[lt]{\lineheight{1.25}\smash{\begin{tabular}[t]{l}$630$\end{tabular}}}}%
    \put(0.60489396,0.0558586){\color[rgb]{0,0,0}\makebox(0,0)[lt]{\lineheight{1.25}\smash{\begin{tabular}[t]{l}$640$\end{tabular}}}}%
    \put(0.80011495,0.0558586){\color[rgb]{0,0,0}\makebox(0,0)[lt]{\lineheight{1.25}\smash{\begin{tabular}[t]{l}$650$\end{tabular}}}}%
  \end{picture}%
\endgroup%